\documentclass[12pt,a4paper]{article}

\usepackage{lmodern}
\usepackage[final]{microtype}
\usepackage[T1]{fontenc}
\usepackage{amssymb}
\usepackage{graphicx}
\usepackage[english]{babel}
\usepackage{braket}
\usepackage{booktabs}
\usepackage{eqnarray}
\usepackage{amsmath}
\usepackage{setspace}
\usepackage{tikz-feynman}
\usepackage[font=small]{caption}
\usepackage{afterpage}
\usepackage[%
backend=biber,%
sorting=none,%
style=numeric-comp,%
maxbibnames=3,%
firstinits=true,%
bibencoding=utf8%
]{biblatex}
\usepackage[final, 
           plainpages=false, 
           pdffitwindow=true,
           bookmarks=true, 
           bookmarksopen=true, 
           bookmarksopenlevel=3, 
           plainpages=false, 
           pdfpagelabels=true, 
           pdfborder={0 0 1}]{hyperref}

\DeclareFieldFormat{pages}{\mkfirstpage{#1}}

\DeclareFieldFormat[article]{journaltitle}{#1}

\AtEveryBibitem{\clearfield{number}}

\renewbibmacro*{name:andothers}{
  \ifboolexpr{
    test {\ifnumequal{\value{listcount}}{\value{liststop}}}
    and
    test \ifmorenames
  }
    {\ifnumgreater{\value{liststop}}{1}
       {\finalandcomma}
       {}%
     \andothersdelim\bibstring[\emph]{andothers}}
    {}}
    
\newbibmacro*{domi:journal}{%
\usebibmacro{journal}%
\setunit{\addspace}%
\mkbibbold{\printfield{volume}}%
\setunit{\addcomma\space}%
\printfield{number}%
\setunit{\addcomma\space}%
\printfield{pages}%
\setunit{\addspace}%
\iffieldundef{year}%
  {}
  {\printtext[parens]{%
  \setunit*{\addspace}%
  \printfield{year}}}%
\nopunct\newunit} 
  
\DeclareBibliographyDriver{article}{%
  \usebibmacro{bibindex}%
  \usebibmacro{begentry}%
  \setunit{\labelnamepunct}\newblock
  \usebibmacro{author} %
  \setunit{\addcomma\space}%
  \usebibmacro{domi:journal}%
  \newunit%
  \usebibmacro{finentry}}
 
\DeclareBibliographyDriver{unpublished}{%
  \usebibmacro{bibindex}%
  \usebibmacro{begentry}%
  \setunit{\labelnamepunct}\newblock
  \usebibmacro{author} %
  \setunit{\addcomma\space}%
  \usebibmacro{title}%
  \setunit{\addcomma\space}%
  \iffieldundef{note}%
    {}
    {\setunit*{\addspace}%
    \printfield{note}}%
  \nopunct\newunit%
  \setunit{\addspace}%
  \iffieldundef{year}%
  {}
  {\printtext[parens]{%
  \setunit*{\addspace}%
  \printfield{year}}}%
  \nopunct
  \usebibmacro{finentry}}
  
\DeclareBibliographyDriver{misc}{%
  \usebibmacro{bibindex}%
  \usebibmacro{begentry}%
  \setunit{\labelnamepunct}\newblock
  \iffieldundef{note}%
    {}
    {\setunit*{\addspace}%
    \printfield{note}}%
  \nopunct
  \iffieldundef{year}%
  {}
  {\printtext[parens]{%
  \setunit*{\addspace}%
  \printfield{year}}}%
  \nopunct
  \usebibmacro{finentry}}
  
\newcommand{\LF}{\mbox{$\Lambda$(1405)} }

\begin{document}
\thispagestyle{empty}

\begin{tabbing}
*************************************************************
\= \kill
\>Exp.-Nr. {\bf A2-07/16  }\\
\>Eingang:           \\
\>an PAC: 31.08.2016    \\\\[-4ex]
\end{tabbing}
\par

\vspace{-4mm}

\begin{center}

{\Large\bf Mainz Microtron MAMI}\\[2ex]
{\bf A2 Collaboration at MAMI} \\
Spokespersons: P.~Pedroni, A.~Thomas \\[2ex]
{\bf Proposal for an Experiment}\\[1ex] 
{\large\bf "Photoproduction of the $\Lambda$(1405) Hyperon"}\\[1ex]
\end{center}
\noindent
{\bf Spokespersons for the Experiment\,:} \\
D.~Werthm\"uller (University of Glasgow, Glasgow, United Kingdom) \\
R.A.~Schumacher (Carnegie Mellon University, Pittsburgh, USA) \\[1ex]
\noindent
{\bf Abstract of Physics\,:}\\
\hspace*{0.5cm}\parbox{16cm}   
{
We propose to use the high current photon beam available at A2
to produce the isosinglet $\Lambda(1405)1/2^-$ hyperon at 
threshold via $\gamma p\rightarrow K^+\Lambda(1405)$.
Its nature is still controversial and actively debated. 
Since the only available photoproduction data from CLAS are lacking in
precision in the $\Sigma^0\pi^0$ decay channel,
we propose a new independent 
measurement of this most important pure $I = 0$ final state including
the unmeasured beam-helicity observable $I^\odot$.
In addition, the
excellent photon detection acceptance of the A2 setup will offer the opportunity
for a first measurement of the radiative decays of the $\Lambda(1405)$,
which will provide clean and stringent constraints for model descriptions in terms
of, e.g., unitary chiral perturbation theory.
} 
\hfill
\\[1ex]
{\bf Abstract of Equipment\,:} \\
\hspace*{0.5cm} \parbox{16cm}          
{
The experiment requires a tagged photon beam above the $\Lambda(1405)$
production threshold using the end-point tagger (EPT).
The charged kaons will be detected using
the in-crystal decay technique in the calorimeters.
Separation of charged pions and protons using the particle identification
detector (PID) and the TAPS vetos/Pizza detector will be important for the
complex and high multiplicity final states. 
The installation of additional detectors (Cerenkov) to improve
the kaon detection efficiency will be investigated.

}\\[1ex]
{\bf MAMI Specifications\,:} \\[-2ex]
\hspace*{0.5cm} \parbox{16cm}
{\begin{tabbing}
******************************\=*********************************\kill
beam energy    \>  1604 MeV     \\
beam polarization  \> polarized  \\     
\end {tabbing}}\\[-4ex]  
{\bf Photon Beam Specifications\,:}\\[-2ex]
\hspace*{0.5cm} \parbox{16cm}
{\begin{tabbing}
******************************\=*********************************\kill
tagged energy range       \>  above 1450 MeV (EPT) \\
photon beam polarization       \>  circularly polarized\\  
\end{tabbing}} \\[-4ex]   
\noindent
{\bf Equipment Specifications\,:} \\[-2ex]
\hspace*{0.5cm} \parbox{16cm}
{\begin{tabbing}
******************************\=************************************\kill
detectors     \>  EPT, CB, TAPS, PID, Pizza detector     \\    
target        \>  liquid hydrogen   \\
\end{tabbing}}\\[-4ex]          
{\bf Beam Time Request\,:} \\[-2ex]
\hspace*{0.5cm} \parbox{16cm}
{\begin{tabbing}
******************************\=***********************************\kill   
set-up/test with beam     \>    150 hours     \\
data taking               \>    1000 hours  \\
\end{tabbing}}

\newpage
{\bf List of participating authors:}

\begin{itemize}

\item {\bf Institut f\"ur Physik, University of Basel, Switzerland}\\
S. Abt, S.~Garni, M.~G\"unther, A.~K{\"a}ser, B.~Krusche, S.~Lutterer, M.~Oberle, Th.~Strub, N.K.~Walford, L.~Witthauer

\vspace{-1ex}
\item {\bf Institut f\"ur Experimentalphysik, University of Bochum, Germany}\\
G.~Reicherz
\vspace{-1ex}

\item {\bf Helmholtz--Institut f\"ur Strahlen- und Kernphysik, University of 
Bonn, Germany}\\
F.~Afzal, R.~Beck, K.~Spieker, A.~Thiel
\vspace{-1ex}


\item {\bf JINR, Dubna, Russia}\\
N.S.~Borisov, A.~Lazarev, A.~Neganov, Yu.A.~Usov 
\vspace{-1ex}

\item {\bf SUPA School of Physics, University of Edinburgh,  UK}\\
M.~Bashkanov, S.~Kay, D.P.~Watts, L.~Zana
\vspace{-1ex}


\item {\bf
SUPA School of Physics and Astronomy, University of Glasgow, UK}\\
J.R.M.~Annand, D.~Hamilton, D.I.~Glazier, S.~Gardner, K.~Livingston, R.Macrae~, I.J.D.~MacGregor, C.~Mullen, D.~Werthm\"uller 
\vspace{-1ex}

\item {\bf Department of Astronomy and Physics, Saint Mary's University Halifax, Canada}\\
A.J.~Sarty
\vspace{-1ex}

\item {\bf Racah Institute of Physics, Hebrew University of Jerusalem, Israel}\\
G.~Ron
\vspace{-1ex}

\item {\bf Kent State University, Kent, USA}\\
C.S.~Akondi, D.M.~Manley
\vspace{-1ex}

\item {\bf Institut f\"ur Kernphysik, University of Mainz, Germany}\\
P.~Achenbach, H.J.~Arends, M.~Biroth, F.~Cividini, A.~Denig, 
P.~Drexler, M.I.~Ferretti-Bondy, W.~Gradl, V.L~Kashevarov, P.P.~Martel,
A.~Neiser, E.~Mornacchi, M.~Ostrick, S.N.~Prakhov, V.~Sokhoyan,
C.~Sfienti, O.~Steffen, M.~Thiel, A.~Thomas, S.~Wagner, J.~Wettig, M.~Wolfes
\vspace{-1ex}

\item {\bf University of Massachusetts, Amherst, USA}\\
R.Miskimen, A.~Rajabi 
\vspace{-1ex}

\item {\bf Institute for Nuclear Research, Moscow, Russia}\\
G.~Gurevic, R.~Kondratiev, V.~Lisin, A.~Polonski
\vspace{-1ex}


\item {\bf INFN Sezione di Pavia, Pavia, Italy}\\
A.~Braghieri, S.~Costanza, P.~Pedroni
\vspace{-1ex}

\item {\bf Department of Physics, Carnegie Mellon University, Pittsburgh, USA}\\
R.A.~Schumacher
\vspace{-1ex}

\item {\bf Department of Physics, University of Regina, Canada}\\
Z.~Ahmed, G.M.~Huber, D.~Paudyal 
\vspace{-1ex}

\item {\bf Mount Allison University, Sackville, Canada}\\ 
D.~Hornidge
\vspace{-1ex}

\item {\bf Tomsk Polytechnic University, Tomsk, Russia}\\ 
A.~Fix
\vspace{-1ex}

\item {\bf George Washington University, Washington, USA}\\
W.J.~Briscoe, C. Collicott, E.J.~Downie, I.I.~Strakovski
\vspace{-1ex}

\item {\bf Rudjer Boskovic Institute, Zagreb, Croatia}\\
M.~Korolija, I.~Supek
\vspace{-1ex}

\end{itemize}


\section{Motivation}

\subsection{Introduction}
The $\Lambda$(1405)$1/2^-$ is a $S = -1$, $I = 0$ hyperon resonance
that lies just below the $\bar{K}N$ threshold 
($E^{thr}_{K^-p}\approx$ 1432 MeV)
and decays via strong interaction to $(\Sigma\pi)^0$ pairs.
The $\bar{K}N$ and the $\Sigma\pi$ channels are coupled via unitarity
and it was first shown in 1959 by Dalitz and Tuan \cite{Dalitz:1959dn}
that the characteristics of the $K^-p$ scattering amplitude 
lead to a resonance in the $\Sigma\pi$ channel.
Soon after this study an excess of events in the $(\Sigma\pi)^0$
system was observed in bubble chamber
experiments measuring 
$K^-p\to \Sigma^+\pi^-\pi^-\pi^+$ and
$K^-p\to \Sigma^-\pi^+\pi^+\pi^-$ \cite{Alston:1961zzd}. 
No excess was found in the
$(\Sigma\pi)^{\pm\pm}$ systems and an isospin of $I = 0$ was deduced 
from the count ratio of the three different $(\Sigma\pi)^0$ observations
using the following isospin representations:
\begin{align}
\ket{\Sigma^+\pi^-} & = \phantom{-} \sqrt{\frac{1}{3}}\ket{0,0} 
                                    + \sqrt{\frac{1}{2}}\ket{1,0}
                                    + \sqrt{\frac{1}{6}}\ket{2,0}
                                    \label{enq:iso_sppm}
                                    \\
\ket{\Sigma^-\pi^+} & = \phantom{-} \sqrt{\frac{1}{3}}\ket{0,0} 
                                    - \sqrt{\frac{1}{2}}\ket{1,0} 
                                    + \sqrt{\frac{1}{6}}\ket{2,0} \\
\ket{\Sigma^0\pi^0} & = -           \sqrt{\frac{1}{3}}\ket{0,0}
                                  \phantom{- \sqrt{\frac{1}{2}}\ket{1,0}}\;
                                    + \sqrt{\frac{2}{3}}\ket{2,0}
                                   \label{eqn:iso_s0pi0}
\end{align}
It should be noted that, neglecting $I = 2$ contributions, 
the $\Sigma^0\pi^0$ system is
a pure $I = 0$ state whereas the charged $\Sigma\pi$ systems have
$I = 1$ contributions as well. This leads for example to the
complication of the $\Sigma$(1385) $I = 1$ resonance contributing
to the charged $\Sigma\pi$ channels. Therefore, the $\Sigma^0\pi^0$
channel provides the cleanest way to study the $\Lambda$(1405).

Today the \LF is listed as a 4-star state in the baryons
listing of the PDG \cite{PDG15}
but despite having the maximum rating, the exact nature of this state
is still not resolved. This is mainly caused by two puzzles.

\begin{figure}[!t]
\centering
\begin{minipage}{.48\textwidth}
\centering
\includegraphics[height=6.5cm]{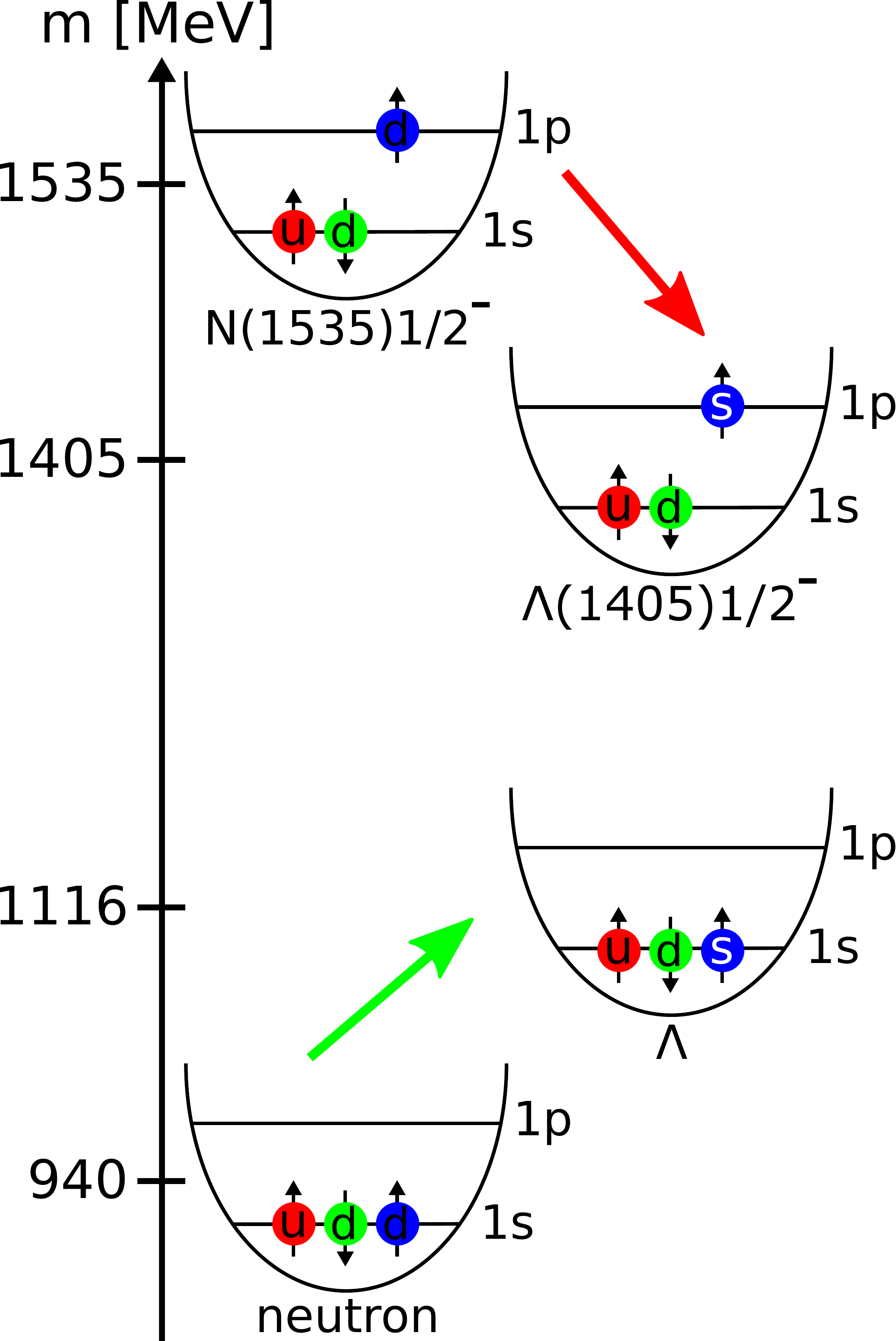}
\caption{Comparison of the ground state and the first $p$-wave
excited state in the simple quark model of the nucleon and 
the $\Lambda$ sector. Despite the $s$-quark, the \LF
is lighter than its $N(1535)$ counterpart.}
\label{fig:levels}
\end{minipage}
\hspace{0.01\textwidth}
\begin{minipage}{.48\textwidth}
\centering
\includegraphics[height=6.5cm]{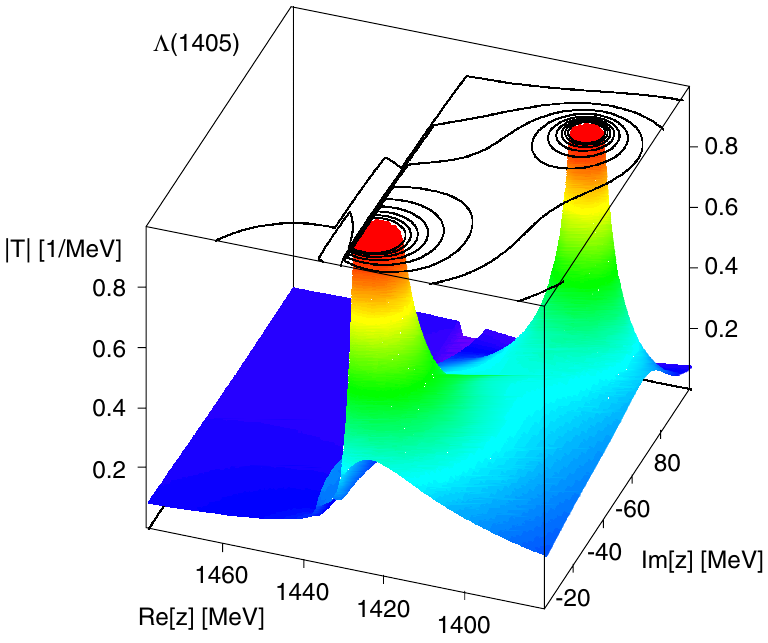}
\caption{Two-pole structure of the \LF resonance in the absolute
value $|T|$ of the elastic $\bar{K}N$ scattering amplitude.
The two poles produce one single broad bump on the real axis.
Taken from \cite{Hyodo:2012df}.}
 \label{fig:poles}
\end{minipage}
\end{figure}

First, most simple three-quark models \cite{Isgur:1978xj,CapstickIsgur:1986} 
fail to describe the low mass of this state when compared to the
situation in the nucleon sector (see Fig.~\ref{fig:levels}). The $p$-wave
excitation of the nucleon ground state $N$(1535)$1/2^-$ generates
a mass difference of about 600 MeV whereas the corresponding 
difference in the $\Lambda$ sector is less than half of this.
In addition, there is a considerable mass gap 
between the spin-orbit partners $\Lambda$(1405)$1/2^-$ and
$\Lambda$(1520)$3/2^-$ in contrast to 
the near-degeneracy of the corresponding nucleon resonances
$N$(1535)$1/2^-$ and $N$(1520)$3/2^-$.
These issues, however, can be coped with by
more sophisticated quark models, e.g.,
a quark-diquark approach in a relativistic quark model inspired
by the heavy quark sector, in which the \LF can be 
naturally reproduced \cite{Faustov:2015eba}.
Alternative descriptions include also exotic configurations such as
compact pentaquarks \cite{Inoue:2007di} and 
hybrids \cite{Kisslinger:2011cx}.

The second enigma about the \LF is the 
claimed two-pole structure 
\cite{Fink:1989uk,Oller:476739,Jido:2003ba,Hyodo:2012df}. 
This is illustrated in 
Fig.~\ref{fig:poles} by the example of the elastic $\bar{K}N$
scattering amplitude.
The two poles in the complex energy plane
both contribute to the amplitude and form a single bump
on the real axis. Therefore, measured observables will always have
contributions from two poles.
The lower wider state around 1325 MeV seems to couple 
more strongly to $\pi\Sigma$, while the higher lying one located
around 1430 MeV close to the $\bar{K}N$ threshold
is more narrow and has a stronger coupling to $\bar{K}N$. 
There are counterarguments to the two-pole scenario 
\cite{Akaishi:2010cr} but it can be shown \cite{Molina:2015tf}
that very recent results of lattice QCD calculations \cite{Hall:2015ej}
seem to support this picture.

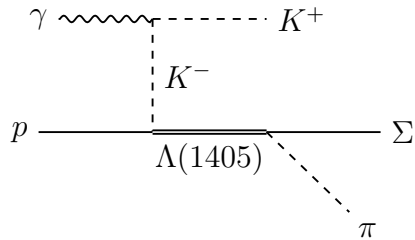
\begin{figure}[b!]
\centering
\begin{tikzpicture}
\begin{feynman}
\vertex (a);
\vertex [left of=a] (i1) {\(\gamma\)};
\vertex [right=of a] (f1) {\(K^+\)} ;
\vertex [below=of a] (b);
\vertex [left=of b] (i2) {\(p\)};
\vertex [right=of b] (c);
\vertex [right=of c] (f2) {\(\Sigma\)};
\vertex [below right=of c] (f3) {\(\pi\)};;
\diagram* {
(i1) -- [photon, line width=0.25mm] (a),
(a) -- [scalar, line width=0.25mm] (f1),
(a) -- [scalar, , line width=0.25mm, edge label=\(K^-\)] (b),
(i2) -- [plain, line width=0.25mm] (b),
(b) -- [double, , line width=0.25mm, edge label'=\(\Lambda(1405)\)] (c),
(c) -- [plain, line width=0.25mm] (f2),
(c) -- [scalar, line width=0.25mm] (f3),
};
\end{feynman}
\end{tikzpicture}
\caption{$K^-$-exchange contribution to $\gamma p\to K^+\Lambda$(1405) 
in the $t$-channel, where the \LF is formed via scattering of the off-shell $K^-$
on the proton.}
\label{fig:feynman}
\end{figure}

Besides experiments with meson beams and bubble chambers
\cite{Thomas:1973uh,Hemingway:1984pz}
the \LF was also studied in photoproduction already in the 1970s
\cite{Boyarski:1970yc} via the $t$-channel dominated reaction
\begin{equation}
\gamma p\to K^+\Lambda(1405)
\end{equation}
As an example, the \LF production via the $K^-$-exchange 
contribution is shown in Fig.~\ref{fig:feynman}. 
Theoretical attention to $\Lambda$(1405)-photoproduction
starting in the late 1990s 
\cite{Nacher:375036,Soyeur:2005bv}
was followed only recently by measurements at LEPS
\cite{Ahn:2003mv,Niiyama:2008iz} and especially at the CLAS experiment 
\cite{Moriya_Lineshape,Moriya_CS,Moriya_JP}, which 
denoted a huge step forward in gaining more experimental data about the 
$\Lambda$(1405). For example, spin and parity were measured directly
for the first time \cite{Moriya_JP} and positions 
and widths of both resonance poles could
be extracted \cite{Roca:2013bq,Roca:2013ev,Mai:2015du}
using chiral unitary theory, which emerged as the 
common framework to describe
dynamics related to the \LF (see \cite{Hyodo:2012df}
for a review).

As mentioned before, theoretical activities concerning 
the \LF are numerous, coming
from fields such as chiral unitary perturbation theory and
lattice QCD. In a very recent work \cite{FernandezRamirez:2016hn}, 
also Regge theory was used
to analyze the latest estimates of
the pole positions suggesting that the higher-lying pole is
consistent with a conventional three-quark picture while the
nature of the lower-lying pole seems to be nonordinary.
This continuing interest from the theory community
motivates thus more experimental attention
to perform further and more precise measurements.

Despite the fact of the CLAS data being of unprecedented quality, 
new measurements could improve on the following points. 
First, the data for the 
most important final state $\pi^0\Sigma^0$ concerning
the study of the $\Lambda$(1405)
were obtained without detection of the three decay photons, leading
to a decrease in statistics due to rigorous cuts in the event selection.
In addition, despite having applied stringent cuts, 
sufficient background rejection
could not be guaranteed \cite{Moriya_Lineshape}.
Second, as delicate experimental parameters, such as detector
acceptance and photon flux, enter directly into the normalized
$m(\Sigma\pi)$ distributions, it is necessary to conduct a 
second measurement at a different experiment
to eliminate effects from those sources of systematic uncertainties. 
Finally, a high statistics measurement very close to threshold would be
desirable as it simplifies the interpretation due to the less complex
situation in this energy region. For example, background
contributions from higher mass hyperon and kaon resonances are not present 
at threshold.

Therefore we propose a new experiment to be performed at A2 that will
provide a significant and independent contribution to the 
experimental database. It will focus on areas where
the A2 experiment has advantages
compared to other experiments.
Namely, we propose to produce the
\LF in photoproduction via $\gamma p\to K^+\Lambda$(1405)
and to measure the following observables:
\begin{enumerate}
\item $m(\Sigma^0\pi^0)$ and $m(\Sigma^+\pi^-)$
      invariant mass distributions (`line shapes')
\item The beam-helicity asymmetry $I^{\odot}$ of 
      $\vec{\gamma} p\to K^+\Sigma^0\pi^0$ and
      $\vec{\gamma} p\to K^+\Sigma^+\pi^-$ 
      with circularly polarized photons
\item Signal search and potential determination of 
      $\Gamma_{Y\gamma} / \Gamma_{\Sigma^0\pi^0}$
      for the $\Lambda$(1405)$\to Y\gamma$ 
      radiative decays with $Y = \{\Lambda, \Sigma^0\}$ 
\end{enumerate}
The next sections will present those physics goals in more detail
discussing previous measurements, the current status of both 
experimental and theoretical research and future contributions
from other experiments.

\subsection[Sigma0pi0 and Sigma+pi- line shapes]{$\Sigma^0\pi^0$ and $\Sigma^+\pi^-$ line shapes}
\label{sec:intro_ls}
Due to the two poles contributing to the $\Lambda$(1405) resonance,
the distribution of the $m(\Sigma\pi)$ invariant mass (`line shape') depends
on the reaction the $\Lambda$(1405) is produced in, giving more weight to 
either the lower or higher lying resonance pole. 
In addition, according to Eqs.~\ref{enq:iso_sppm}--\ref{eqn:iso_s0pi0},
the different $(\Sigma\pi)^0$
isospin states are sensitive to the isospin structure of the 
production amplitude.

\begin{figure}[!t]
\centering
\begin{minipage}{.48\textwidth}
\centering
\includegraphics[height=6.15cm]{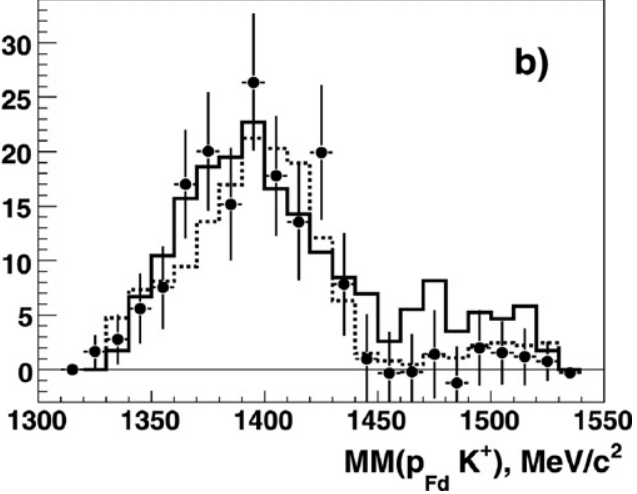}
\caption{ANKE $pp\to pK^+\Sigma^0\pi^0$ results (points)
compared to $\pi^-p\to K^0 (\Sigma\pi)^0$ \cite{Thomas:1973uh}
(solid line)
and $K^-p\to\pi^+\pi^-\Sigma^+\pi^-$ \cite{Hemingway:1984pz}
(dotted line).}
\label{fig:ls_anke}
\end{minipage}
\hspace{0.01\textwidth}
\begin{minipage}{.48\textwidth}
\centering
\includegraphics[height=6.5cm]{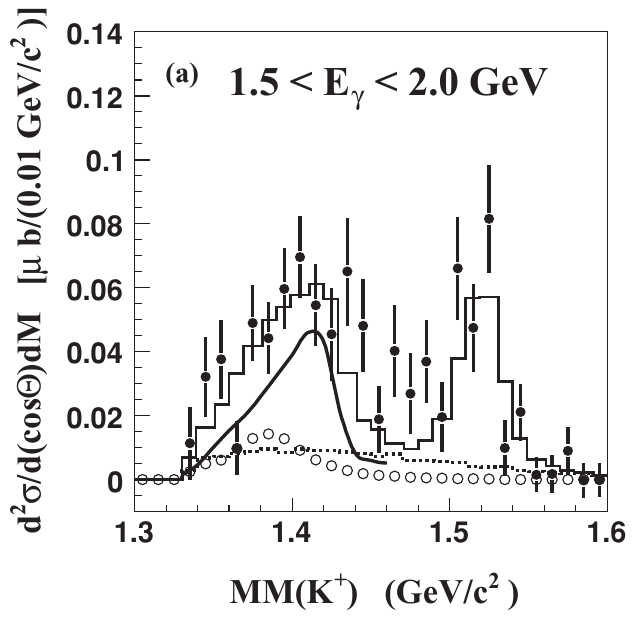}
\caption{LEPS $K^+$ missing mass of \mbox{$\gamma p\to K^+\Lambda$(1405)}, 
$\Lambda$(1405)$\to\Sigma^\pm\pi^\mp$ event candidates
\cite{Niiyama:2008iz}. The theoretical line shape from 
\cite{Nacher:375036} was fitted along with other contributions
to the data.}
\label{fig:ls_leps}
\end{minipage}
\end{figure}

The $\Sigma^0\pi^0$ channel
bears the advantage that only isospin $I = 0$ amplitudes contribute
(apart from negligible non-resonant $I = 2$ terms, 
see Eq.~\ref{eqn:iso_s0pi0}), which
facilitates the theoretical interpretation of experimental data, 
while the nearby $\Sigma^0$(1385) resonance decaying into $\Sigma^+\pi^-$
and $\Sigma^-\pi^+$ makes an isolated study of the $\Lambda$(1405) 
in those decay channels more difficult. 
Unfortunately, most older experiments were not able to
measure the neutral pion and the decay photon from 
the $\Sigma^0\to\Lambda\gamma$
decay in the $\Sigma^0\pi^0$ final state.
For example, older bubble chambers experiments 
using $\pi^-$ \cite{Thomas:1973uh} and $K^-$ beams
\cite{Hemingway:1984pz} were only able to extract
the line shapes of the $\Sigma^\pm\pi^\mp$ final states.

The Crystal Ball collaboration was able to show
the presence of the \LF in the $\Sigma^0\pi^0$ system
in the measurement of $K^-p\to\pi^0\pi^0\Sigma^0$
\cite{Prakhov:2004iw}. However, due to the two
indistinguishable $\pi^0$-meson this reaction is 
less suited to study the \LF in
the $\Sigma^0\pi^0$ distribution.

More recently, the $(\Sigma\pi)^0$ line shapes were also
studied in $pp$ collisions at the ANKE \cite{Zychor:2008gt}
and HADES \cite{Agakishiev:2013ba} experiments. While
the result of the $\Sigma^0\pi^0$ channel obtained at
ANKE were found to be in reasonable agreement with
the older meson-beam bubble chamber data 
(see Fig.~\ref{fig:ls_anke}),
the $\Sigma^\pm\pi^\mp$ distributions measured by HADES 
differ from those older results. This illustrates the 
influence of the different production mechanisms and 
final states.

\begin{figure}[!t]
\centering
\includegraphics[width=9cm]{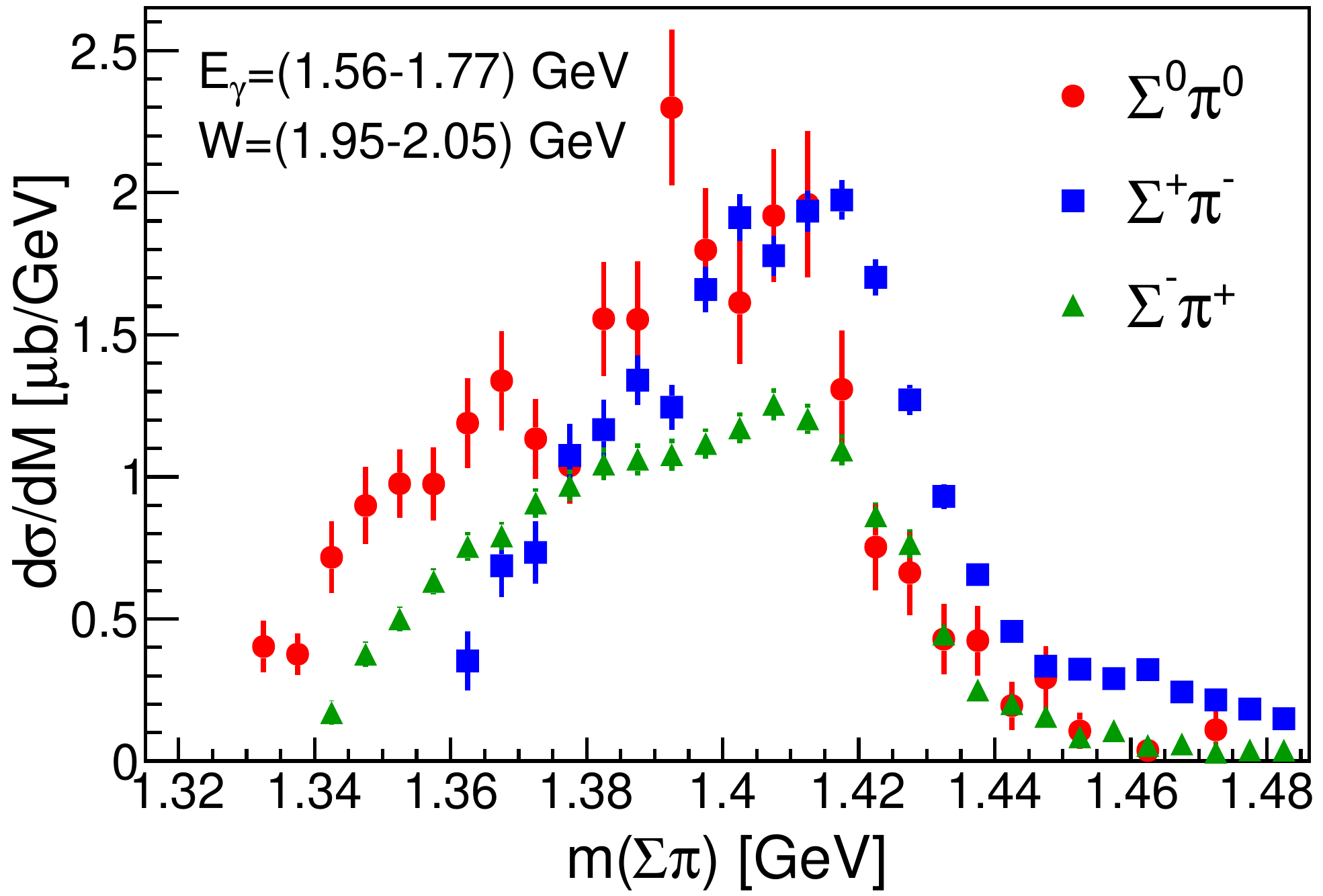}
\caption{$(\Sigma\pi)^0$ line shapes measured by the CLAS collaboration
\cite{Moriya_Lineshape}
via $\gamma p\to K^+(\Sigma\pi)^0$
in the photon energy bin closest to threshold. The error bars
represent only the statistical uncertainties.}
\label{fig:ls_clas}
\end{figure}

The line shapes in the $\Sigma^\pm\pi^\mp$ channels
were also studied in photoproduction experiments at LEPS
\cite{Ahn:2003mv,Niiyama:2008iz}. The result of the lower 
photon energy bin shown in Fig.~\ref{fig:ls_leps} shows
good agreement with the theoretical calculation by Nacher
{\it et al.} \cite{Nacher:375036}.
Finally, the CLAS collaboration obtained high statistics
data from photoproduction measuring all three $(\Sigma\pi)^0$
isospin channels \cite{Moriya_Lineshape}. The line shapes
closest to threshold and the kinematic region accessible
with the proposed A2 experiment are shown in Fig.~\ref{fig:ls_clas}.
As mentioned before, the $\Sigma^0\pi^0$ invariant mass
distribution suffers from higher systematic and statistical 
uncertainties compared to the $\Sigma^\pm\pi^\mp$ channels.
This is caused by the very limited to nonexistent photon
detection efficiency of the CLAS detector, which was optimized
for the detection of charged particles. In this respect the
A2 experiment provides a completely complementary experimental
setup being optimized for photon detection in almost the
complete solid angle with still acceptable capabilities for the
detection of charged particles.

The CLAS photoproduction data were found to be valuable input for
unitary chiral approaches in terms of extracting the two
\LF pole positions \cite{Roca:2013bq,Roca:2013ev} and
eliminating solutions that describe the existing
hadronic data equally well \cite{Mai:2015du}. Therefore, new 
data of the proposed experiment could be easily analyzed 
by those groups using existing frameworks.

\subsection[Beam-helicity asymmetry Icirc of 
gp->KPlus(Sigma pi)0]
{Beam-helicity asymmetry $I^{\odot}$ of 
$\vec{\gamma} p\to K^+(\Sigma\pi)^0$}
\label{sec:intro_Icirc}
The beam-helicity asymmetry $I^{\odot}$ is a polarization 
observable in photoproduction defined for three-body 
final states, e.g. $N\pi\pi$ \cite{Roberts:2005hs}, that can be 
measured with a circularly polarized beam and an unpolarized 
target. It is defined as
\begin{equation}
I^\odot(\Phi) = \frac{d\sigma^+ - d\sigma^-}{d\sigma^+ + d\sigma^-} =
\frac{1}{P_\gamma} \frac{N^+ - N^-}{N^+ + N^-},
\label{eqn:i_circ}
\end{equation}
with the differential cross sections $d\sigma^\pm$ for the
two photon helicity states, the degree of circular polarization
of the photon $P_\gamma$ and an angle $\Phi$ that can be defined
in various ways in the center-of-mass system of the photon
and the initial state nucleon. For $\gamma p\to p \pi\pi$
sensitivity to small contributions via interference terms
was found \cite{Roca:2005ic}.
In addition, the
experimental access to this observable is particularly straightforward
since all normalization factors cancel in the ratio, apart from
detection efficiencies when integrated over angles or photon 
beam energies. Therefore, the experimental extraction only relies
on the observed counts $N^\pm(\Phi)$ for the two beam helicity states.

In double-pion photoproduction $I^\odot$  has been measured with the A2 
experiment for several reactions 
\cite{Krambrich:2009ec,Oberle:2013ks,Oberle:2014kb}.
The degree of longitudinal polarization of the
electrons at MAMI is usually $P_{e^-}\approx$ 70--85\% and
is transferred to the circular polarization of the photons
according to Olsen and Maximon \cite{Olsen:1959lf}. In the energy range
covered by the endpoint tagger for the proposed experiment,
this basically translates into a constant values for $P_\gamma$
that is very close to $P_{e^-}$.

There are no measurements known to the authors of the 
observable $I^\odot$
for the reactions $\vec{\gamma} p\to K^+(\Sigma\pi)^0$.
On the theory side, calculations exists for
$\gamma N\to K\bar{K}N$, where sensitivity of
certain observables to the \LF coupling to $\bar{K}N$ 
was found and measurements of $K\Sigma\pi$ final states
were encouraged \cite{Roberts:2006gz}.

\begin{figure}[!t]
\centering
\begin{minipage}{.40\textwidth}
\centering
\includegraphics[width=\textwidth]{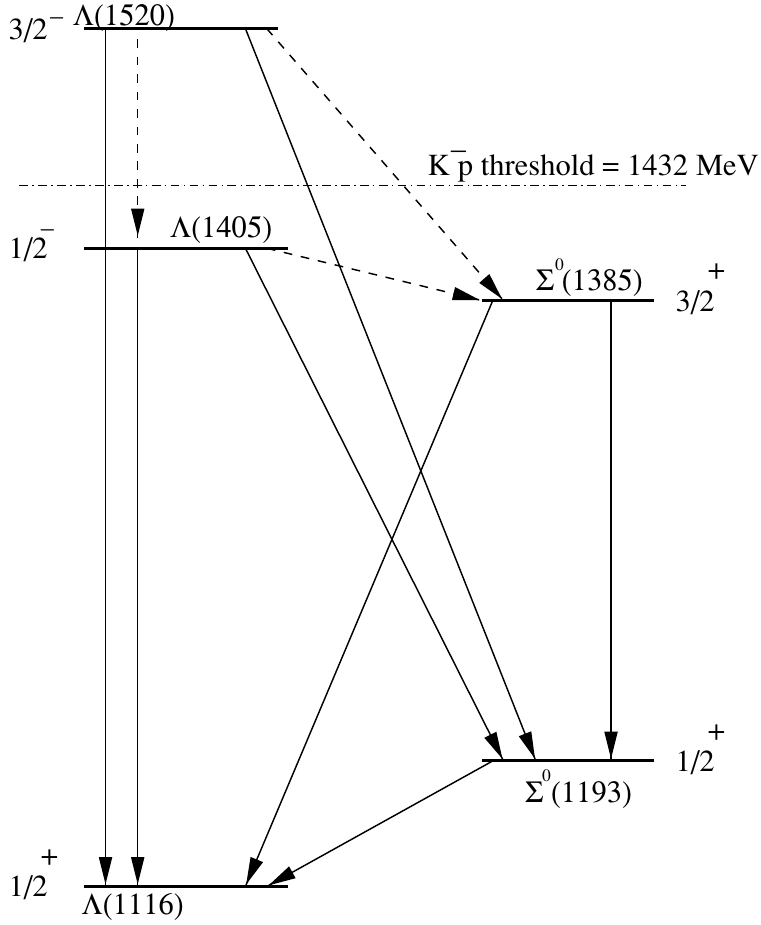}
\caption{
Radiative decay scheme of low-lying hyperon states. Suppressed
transitions are denoted by the dashed lines.
Taken from \cite{Taylor:2005bn}.
}
\label{fig:hyperon_decays}
\end{minipage}
\hspace{0.01\textwidth}
\begin{minipage}{.57\textwidth}
\centering
\vspace{3mm}
\includegraphics[width=\textwidth]{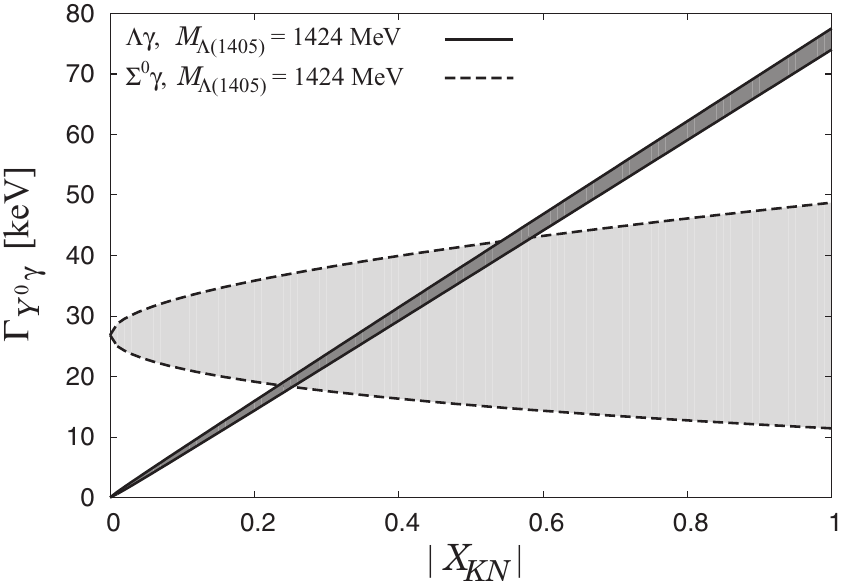}
\vspace{2mm}
\caption{
Radiative decay widths of the \LF as functions of the absolute
value of the $\bar{K}N$ compositeness for $m_{\Lambda(1405)} = 1424$ MeV 
\cite{Sekihara:2014dq}.
}
\label{fig:l1405_rad_comp}
\end{minipage}
\end{figure}

\subsection[Lambda(1405) radiative decays]{\LF radiative decays}
\label{sec:intro_raddec}
A clean way of probing the structure of baryons is the
study of their radiative decays by emission of a photon. The decay scheme
for the low-lying hyperon states is shown in Fig.~\ref{fig:hyperon_decays}.
For the \LF an early calculation
\cite{Darewych:1983yw} within the nonrelativistic quark model \cite{Isgur:1978xj}
obtained
\begin{align*}
\Gamma_{\Lambda\gamma}  & =  143\; \mathrm{keV} &
                        \Gamma_{\Lambda\gamma} / \Gamma & = 2.83 \times 10^{-3} \\
\Gamma_{\Sigma^0\gamma} & =  91\; \mathrm{keV} &
                        \Gamma_{\Sigma^0\gamma} / \Gamma & = 1.80 \times 10^{-3}
\end{align*}
with $\Gamma = (50.5 \pm 2.0)$ MeV \cite{PDG15}.
These values are very small and pose a challenge to experiments --- yet today
there are still no experimental data available that were directly measured.
There is a model-dependent determination 
\cite{Burkhardt:1991ms} of the radiative widths extracted from measurements 
of the decays of kaonic hydrogen \cite{Whitehouse:1989bs}
which found even smaller values:
\begin{align*}
\Gamma_{\Lambda\gamma}  & =  (27 \pm 8) \; \mathrm{keV} &
                            \Gamma_{\Lambda\gamma} / \Gamma & = (5.35 \pm 1.60) \times 10^{-4} \\
\Gamma_{\Sigma^0\gamma, 1} & =  (10 \pm 4) \; \mathrm{keV} &
                             \Gamma_{\Sigma^0\gamma, 1} / \Gamma & 
                                              = (1.98 \pm 0.80) \times 10^{-4}\;\;\mathrm{or} \\
\Gamma_{\Sigma^0\gamma, 2} & =  (23 \pm 7) \; \mathrm{keV} &
                             \Gamma_{\Sigma^0\gamma, 2} / \Gamma & 
                                              = (4.55 \pm 1.40) \times 10^{-4}.
\end{align*}
Several models
describing the $\Lambda$(1405) predict very different values for the two
decay widths and in particular for their ratio. Namely, by using the
two-pole results from chiral unitary theories, it was found that the
ratio $\Gamma_{\Lambda\gamma} / \Gamma_{\Sigma^0\gamma}$ is reversed
for the lower and the higher-lying pole \cite{Geng:2007fv}.
Hence, experimental data on the radiative decay widths provide an
excellent test of the conjectured two-pole structure or even
speculative five-quark components \cite{An:2010fx}. In another work
\cite{Sekihara:2014dq}, a relation between the compositeness, which is a measure
for the amount of the $\bar{K}N$ component inside the $\Lambda$(1405), and
the radiative decay width is established. The relation for both radiative decays
is shown in Fig.~\ref{fig:l1405_rad_comp} evaluated for a \LF mass of 1424 MeV.
This value corresponds to the higher-lying pole, where the relation is more
striking. Due to cancelling $\Sigma^\pm\pi^\mp$ terms
in the $\Lambda\gamma$ decay mode, the $K^-p$ component dominates this decay
resulting in a linear relationship between $\Gamma_{\Lambda\gamma}$ and the
$\bar{K}N$ compositeness. Using this calculation and a direct measurement of 
$\Gamma_{\Lambda\gamma}$ from photoproduction, the meson-baryon component
in the \LF could thus be estimated. 

The authors of \cite{Sekihara:2014dq} also conduct a reevaluation of the 
\LF radiative width in the chiral unitary approach with
updated parameters and obtain the following values:
\begin{align}
\label{unit_chi_pol_res_lg}
\Gamma_{\Lambda\gamma}   & =  31\; \mathrm{keV\ (lower\ pole)} &
\Gamma_{\Lambda\gamma}   & =  96\; \mathrm{keV\ (higher\ pole)} \\
\label{unit_chi_pol_res_sg}
\Gamma_{\Sigma^0\gamma}  & =  94\; \mathrm{keV\ (lower\ pole)} &
\Gamma_{\Sigma^0\gamma}  & =  60\; \mathrm{keV\ (higher\ pole)}
\end{align}
In view of all those possibilities on the theoretical side, experimental
input from a direct measurement of the radiative decay widths
is urgently needed.
More recent attempts to measure the radiative decays of the $\Lambda$(1405) in the 
CLAS experiment \cite{Taylor:2005bn,Keller:2011br} failed due to contamination from the
overlapping $\Sigma^0$(1385) resonance and the lack of detecting 
the radiated photon. 
Here again, the A2 setup with its excellent photon 
detection capabilities and high solid angle coverage could provide
a great opportunity for a new attempt to observe and quantity the
radiative decays of the $\Lambda$(1405).

\subsection{Competing experiments}
There is an ongoing interest to study the \LF at several 
experiments worldwide. Information about the low energy
$K^-p$ interaction was obtained in the SIDDHARTA experiment
at DA$\Phi$NE \cite{Bazzi:2011cv} by studying kaonic hydrogen.
An upgrade to this experiment measuring the $K^-d$ interaction
seems to be planned \cite{Mai:2015du}. The AMADEUS experiment,
which was part of the KLOE detector located at the same facility,
published prelimary results of the $\Sigma^0\pi^0$ line shape
\cite{Piscicchia:2013df}.

At J-PARC there are continuing activities related to
strangeness physics and the \LF
at various experiments (E31/E42/E45) using $\pi^-$ and $K^-$
beams. For example, the E31 collaboration is working on
$(\Sigma\pi)^0$ line shape results \cite{Kawasaki_Menu16}
and the E45 collaboration with the 
HypTPC detector is planning measurements of the \LF
radiative decay \cite{Sako:PWA8}. As the setup seems to
be incapable of photon detection, the experiment will
rely on good momentum resolution of the charged particles
to use the missing energy technique for an indirect
measurement of the radiative decay.

Competing photoproduction experiments using
beam energies up to $\sim$3 GeV will be
LEPS2 at SPring-8 \cite{Nakano_Menu16}, especially once 
the new spectrometer based on
the BNL-E949 solenoid will be operational, and the
BGO-OD experiment at ELSA \cite{Jude_NSTAR15_Proc}.
The latter was specifically built for the study
of hyperon resonances with excellent kaon detection
capabilities in forward direction although the detection
is restricted to quite a small solid angle. Also,
photons cannot be detected in this region and the
available photon beam intensity is lower compared to
MAMI. In addition, simulations 
(see Sec.~\ref{sec:kin_rec}) have shown that
a considerable amount of kaons and protons coming from
the reactions of interest are going to regions where
they cannot be detected or not very precisely measured
in this experiment.

Finally, photoproduction experiments with beam energies
above 3 GeV will be the next generation experiments 
GlueX and CLAS12
at the Thomas Jefferson National Accelerator Facility (TJNAF).
At CLAS12, quasi-real tagged photons will be available via the
forward tagger with energies normally around 6.5--10.5 GeV.
The very high momentum kaons in forward direction originating
from the $t$-channel production mechanism may be detected in CLAS12
but the decay photons going beyond 35 degrees in lab polar angle will
not be detected \cite{CLAS12_Tech_Design_Rep}. This will result in
a low detection efficiency and higher background contamination.
The hermetic GlueX 
detector has good photon
detection capabilities over a large part of the
solid angle but the discrimination
of high energetic kaons will only be available in 
2018 \cite{StevensDirc:2016gd}.

\section{Proposed experiment}
The proposed experiment aims at measuring
$\vec{\gamma} p\to K^+\Lambda$(1405) 
with a circularly polarized photon beam
and an unpolarized liquid hydrogen (LH$_2$) target
to achieve the physics goals presented in
Secs.~\ref{sec:intro_ls}--\ref{sec:intro_raddec}.
This will lead to the following primary final states:
\setstretch{1.2}
\\[0.5\baselineskip]
\begin{tabular}{@{}lllll@{}}
(R1) &
\LF$\to\Sigma^0\pi^0$ & 
$\Sigma^0\to\Lambda\gamma$ &  
$\Lambda\to p\pi^-$ &
$\to K^+ p\pi^- 3\gamma$ \\
(R2) &
\LF$\to\Sigma^+\pi^-$ & 
$\Sigma^+\to p\pi^0$ &  
&
$\to K^+ p\pi^- 2\gamma$ \\
(R3) &
\LF$\to\Lambda\gamma$ & 
&  
$\Lambda\to p\pi^-$ &
$\to K^+ p\pi^- \gamma$ \\
(R4) &
\LF$\to\Sigma^0\gamma$ & 
$\Sigma^0\to\Lambda\gamma$ & 
$\Lambda\to p\pi^-$ &
$\to K^+ p\pi^- 2\gamma$ \\
\end{tabular}
\\[0.5\baselineskip]
\setstretch{1}
It can be seen that two final states are identical and the others
only differ in the number of photons. Therefore, fully exclusive
measurements are inevitable to discriminate the different event 
candidates and a high angular coverage for photons, as 
provided by the A2 setup, is essential to minimize
event contamination caused by undetected particles, 
especially photons.

The A2 setup offers the possibility to exploit the
$\Lambda\to n\pi^0$ decay in addition to
$\Lambda\to p\pi^-$ as done before
using the Crystal Ball detector \cite{Prakhov:2004iw}.
This will lead to alternative
event samples for reactions (R1)--(R4), albeit with lower 
statistics due to the smaller branching ratio of the 
neutral $\Lambda$ decay, the lower detection efficiency for neutrons
and the higher cluster multiplicity. Nevertheless, these additional
events could be useful for cross checks.

On the other hand, it will most certainly not
be possible to measure 
the second charged \LF$\to\Sigma^-\pi^+$ decay since the
$\Sigma^-\to n\pi^-$ leads to the two most 
problematic particles for the A2 setup. The energy of neutrons
cannot be measured and the reconstructed energies of $\pi^-$-mesons
have large uncertainties. One of those particles in the final state
can be handled by calculating its energy from kinematics
and the energies of the other particles. This will not be possible
if there are two of them.

\begin{table}[!t]
\centering
\begin{tabular}{@{}lcccccc@{}}
\toprule
& $\Lambda(1405)$ & $\Sigma^0(1385)$ & $\Lambda$ & $\Sigma^0$ & $\Sigma^+$ & $\Sigma^-$\\
\midrule
$\Sigma^+\pi^-$  & 33\%$^1$ & 5.9\%$^1$   & ---  & ---   & --- & --- \\
$\Sigma^-\pi^+$  & 33\%$^1$ & 5.9\%$^1$   & ---  & ---   & --- & --- \\	
$\Sigma^0\pi^0$  & 33\%$^1$ & ---         & ---  & ---   & --- & --- \\
$\Lambda\pi^0$   & ---      & 87\%        & ---  & ---   & --- & --- \\[2mm]
$\Lambda\gamma$  &  ?       & 1.25\%      & ---  & 100\% & --- & --- \\
$\Sigma^0\gamma$ & ?        & ?           & ---  & ---   & --- & --- \\[2mm]
$p\pi^-$         & ---      & ---         & 64\% & ---   & --- & --- \\
$n\pi^0$         & ---      & ---         & 36\% & ---   & ---   & --- \\
$p\pi^0$         & ---      & ---         & ---  & ---   & 52\%  & --- \\
$n\pi^+$         & ---      & ---         & ---  & ---   & 48\%  & --- \\
$n\pi^-$         & ---      & ---         & ---  & ---   & ---   & 100\% \\
\bottomrule
\end{tabular}
\caption{Decay branching ratios $\Gamma_i / \Gamma$ of the involved
hyperons. $^1$assuming isospin symmetry.
All values from \cite{PDG15}.}
\label{tab:hyperon_decays}
\end{table}

Tab.~\ref{tab:hyperon_decays} gives an overview of the decays
and the corresponding branching ratios of all involved hyperons.

\subsection{Experimental setup and particle detection}
The experiment will be performed using available A2
equipment. Due to the high \LF production threshold 
of $E_\gamma\approx 1450$ MeV the electron beam delivered
by MAMI should have the maximum energy of 1604 MeV.
In this configuration, the standard Glasgow photon tagger
\cite{McGeorge_08}
can only tag photons up to 
$\sim$1490 MeV. This leads to a tagged 
range in the region of interest of only 40 MeV, where the production 
cross section is still very small. The so-called endpoint tagger (EPT) 
\cite{Adlarson:2015im} was
specifically constructed to cover the very high photon energy region
of the A2 bremsstrahlung spectrum. It can be configured to cover
photon energies from 1450--1590 MeV with a resolution of 2.7--3 MeV
\cite{AhrensPC:2016dg}. Parallel running along with the main tagger is
not possible, therefore experiments running with the EPT require 
a modification of the A2 beamline. During 2014, a series of experiments
dedicated to the study of $\eta'$-decays have taken place
\cite{Unverz_Prop}, in which the EPT was used. Parts of the data 
obtained in these measurements have been analyzed for the preparation
of this proposal and were found to be very useful for the determination
of the running conditions for the proposed experiment 
(see Sec.~\ref{sec:exp_run_cond}).

\begin{figure}[t]
\centering
\includegraphics[width=8cm]{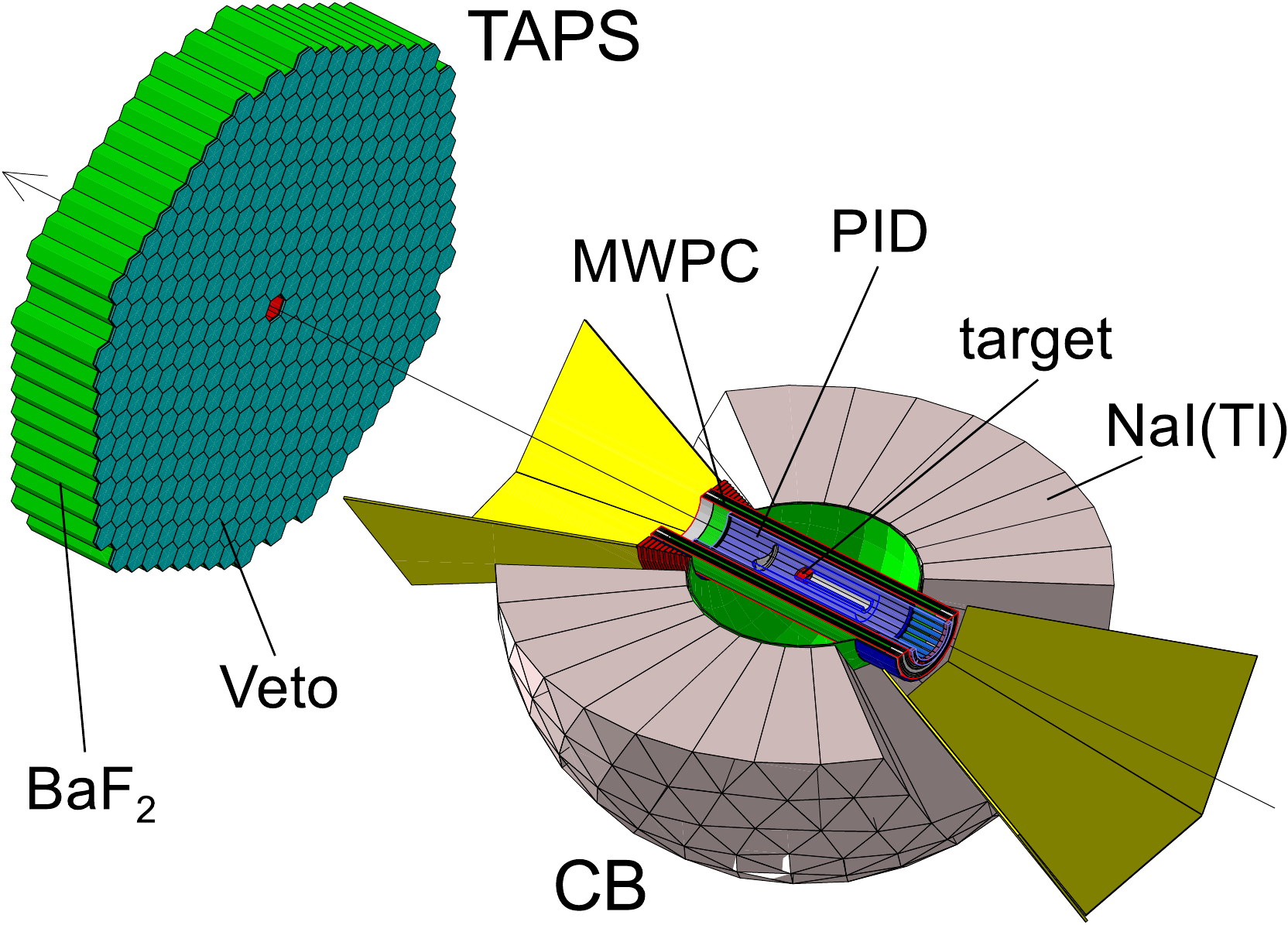}
\caption{Schematic view of the A2 experimental setup with the
central detector Crystal Ball (CB) and the forward calorimeter TAPS, and
the corresponding subcomponents for charged particle discrimination (PID, Veto)
and tracking (MWPC).}
\label{fig:exp_setup}
\end{figure}

The main detector setup is shown schematically in Fig.~\ref{fig:exp_setup}.
The bremsstrahlung photons will impinge on the liquid hydrogen
target installed in the center of the Crystal Ball (CB) 
detector \cite{Starostin_01}. It consists of two hemispheres with in total 672 
optically insulated NaI(Tl) crystals of 15.7 radiation length thickness, covering all 
azimuthal angles for the polar angle range $20^\circ < \theta < 160^\circ$. 
All crystals point towards the center of the sphere.
The distance from the center to the detector modules is 25 cm.
The energy resolution for photons can be described as 
$\Delta E / E = 2\%/(E[\mathrm{GeV}])^{0.36}$ while typical angular resolutions are 
$\Delta\theta = 2^\circ$--$3^\circ$ and $\Delta\phi = 2^\circ$--$4^\circ$ 
\cite{Werthmuller:2014thb}. The multi-wire proportional chamber (MWPC) 
surrounding the PID provides basic tracking and can be used improve the angular
resolution of charged particles in CB.

The forward hole of CB is covered by the hexagonal TAPS wall, which is made of 
366 hexagonally shaped BaF$_2$ crystals with a thickness of 12 radiation lengths
and an inner ring of 72 PbWO$_4$ crystals of 22.5 radiation lengths
at small forward angles.
TAPS is normally installed 1.46 m downstream from the target covering polar angles from 
$5^\circ$ to $21^\circ$. The photon energy resolution is parametrized as 
$\Delta E / E = 1.8\% + 0.8\%/(E[\mathrm{GeV}])^{0.5}$ \cite{Gabler_94}. 
The fine granularity of the detector
elements leads to excellent resolution in the polar angle (better than $1^\circ$), while 
$\Delta\phi = 1^\circ$--$6^\circ$. Neutral and charged particles 
can be discriminated by plastic scintillators in 
both detectors. A 50 cm long barrel of 24 strips with a width of 4 mm surrounds 
the target and acts as particle identification detector (PID) for 
CB \cite{Watts_05}. In TAPS charged particles 
can be identified with individual 5 mm thick plastic scintillators that are installed
in front of every detector element. As the $dE/E$-resolution is worse compared
to the PID, alternatively the recently constructed pizza detector could be used.
It consists of 24 plastic scintillator sectors and
would be place in front of TAPS. Tests measurements are needed to
determine the achievable $dE/E$-resolution but as the scintillator
thickness is 1 cm, it is expected to be rather good \cite{AhrensPC:2016dg}.
TAPS additionally provides separation of photons and massive particles
via time-of-flight measurements and by
using the two scintillation light components in the BaF$_2$ crystals
in a pulse-shape analysis.

Although the A2 setup was optimized for the detection of photons,
neutrons and charged particles can be detected as well with
the following restrictions:

\paragraph{Neutrons} 
Neutrons can be detected in both CB and TAPS with typical efficiencies
up to 40\%. Separation from photons is only possible in TAPS (time-of-flight,
BaF$_2$ pulse-shape analysis), separation from protons both in TAPS and CB. An
energy measurement is not possible since the kinetic energy cannot
be deduced from the deposited energy \cite{Werthmuller:2014thb}.

\paragraph{Protons}
Protons can be detected in both CB and TAPS with typical efficiencies around 80\%.
The kinetic energy can be determined from the deposited energy using
appropriate corrections up to the punch-through energy (CB: $\sim$400 MeV, 
TAPS: $\sim$380 MeV). Discrimination to pions is possible using the
$dE/E$-technique in CB (up to 250 MeV) and TAPS (up to 150 MeV).

\paragraph{Pions}
Pions can be detected in both CB and TAPS with typical efficiencies up to 70\%.
The kinetic energy can be determined 
from the deposited energy using
appropriate corrections up to the punch-through energy (CB: $\sim$250 MeV, 
TAPS: $\sim$200 MeV). The energy resolution for $\pi^-$-mesons is worse since,
once at rest, they will form pionic atoms, which will finally lead
to additional energy depositions by emitted protons and photons
\cite{Marin:1998zf}. Discrimination of charged pions to protons 
is possible using the $dE/E$-technique in CB (up to 250 MeV) 
and TAPS (up to 150 MeV) but the charge of the pions cannot be
determined due to the absence of a magnetic field in the setup.

\paragraph{Kaons}
$K^+$-mesons can be detected with the in-crystal decay technique
\cite{Jude_KP} in both CB and TAPS with typical efficiencies of 20--30\%.
If the kinetic energy of the kaons is below the punch-through energy
(CB: $\sim$350 MeV, TAPS: $\sim$330 MeV), they will be stopped in the
crystals and decay after a mean lifetime of 12.38 ns
\cite{PDG15}. The decay products will deposit additional energy
in a secondary cluster connected to the main cluster.
An algorithm was developed to search for
such cluster signatures \cite{Jude_KP}. It splits potential $K^+$-clusters
into impact and decay subclusters based on 
the timing signals. Energy
and direction of the kaon can then be accessed
via the impact cluster, while properties of the
decay cluster help to differentiate between
the dominant
$\mu^+\nu_\mu$ ($\Gamma_i/\Gamma = 63.56\%$) and
$\pi^+\pi^0$ ($\Gamma_i/\Gamma = 20.67\%$) \cite{PDG15}
decays.

\begin{figure}[t]
\centering
\includegraphics[width=0.32\textwidth]{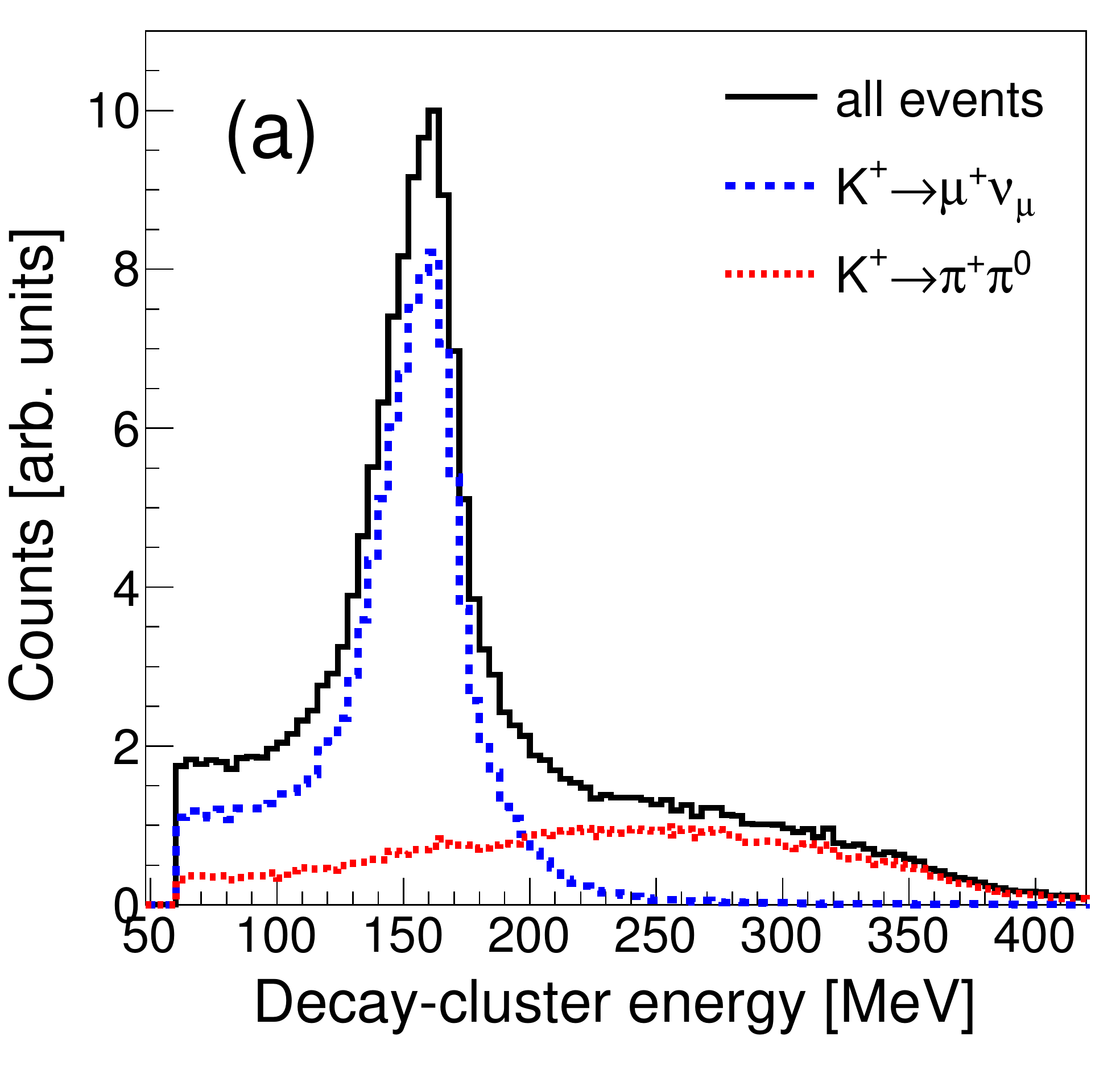}
\includegraphics[width=0.32\textwidth]{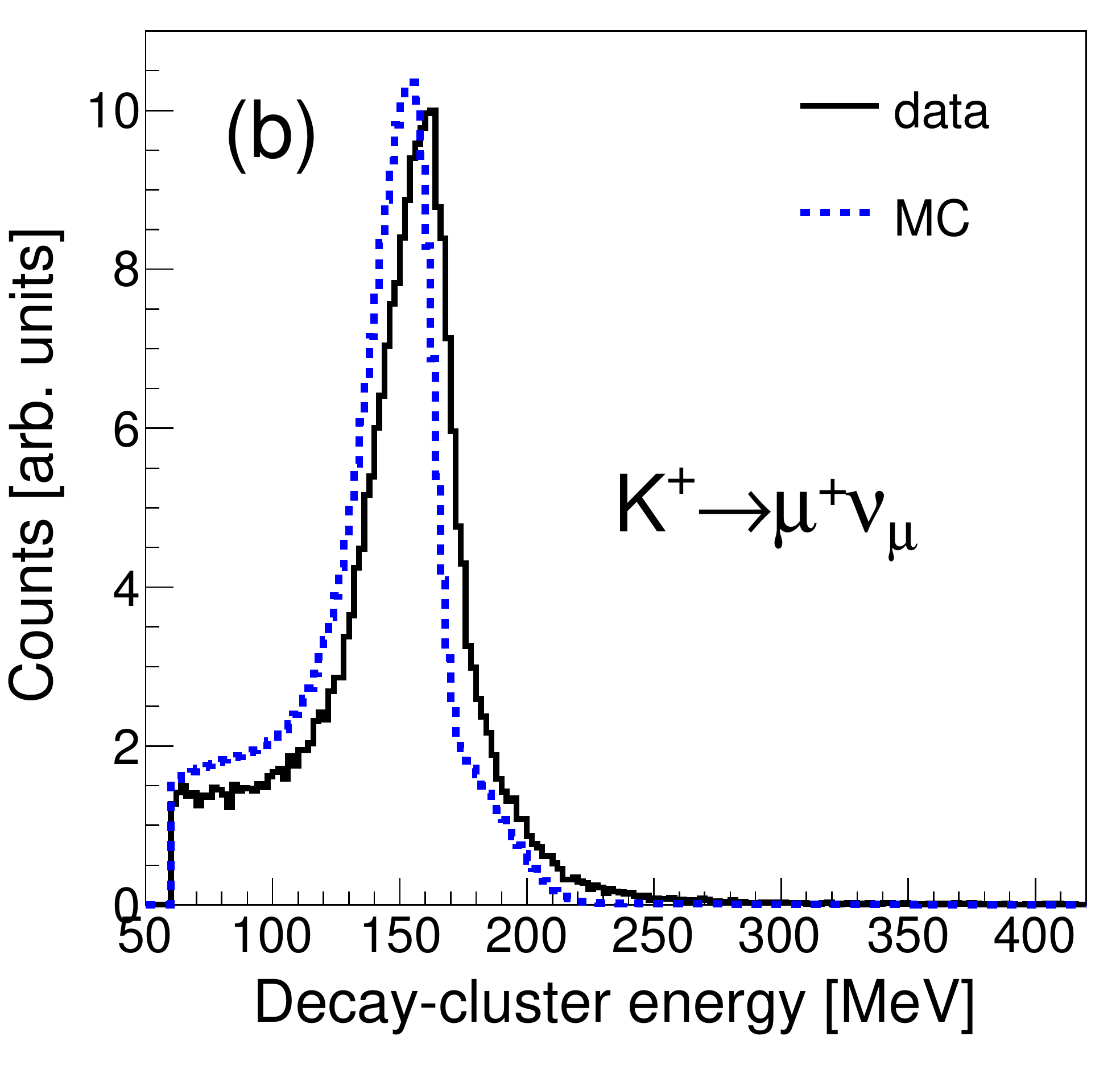}
\includegraphics[width=0.32\textwidth]{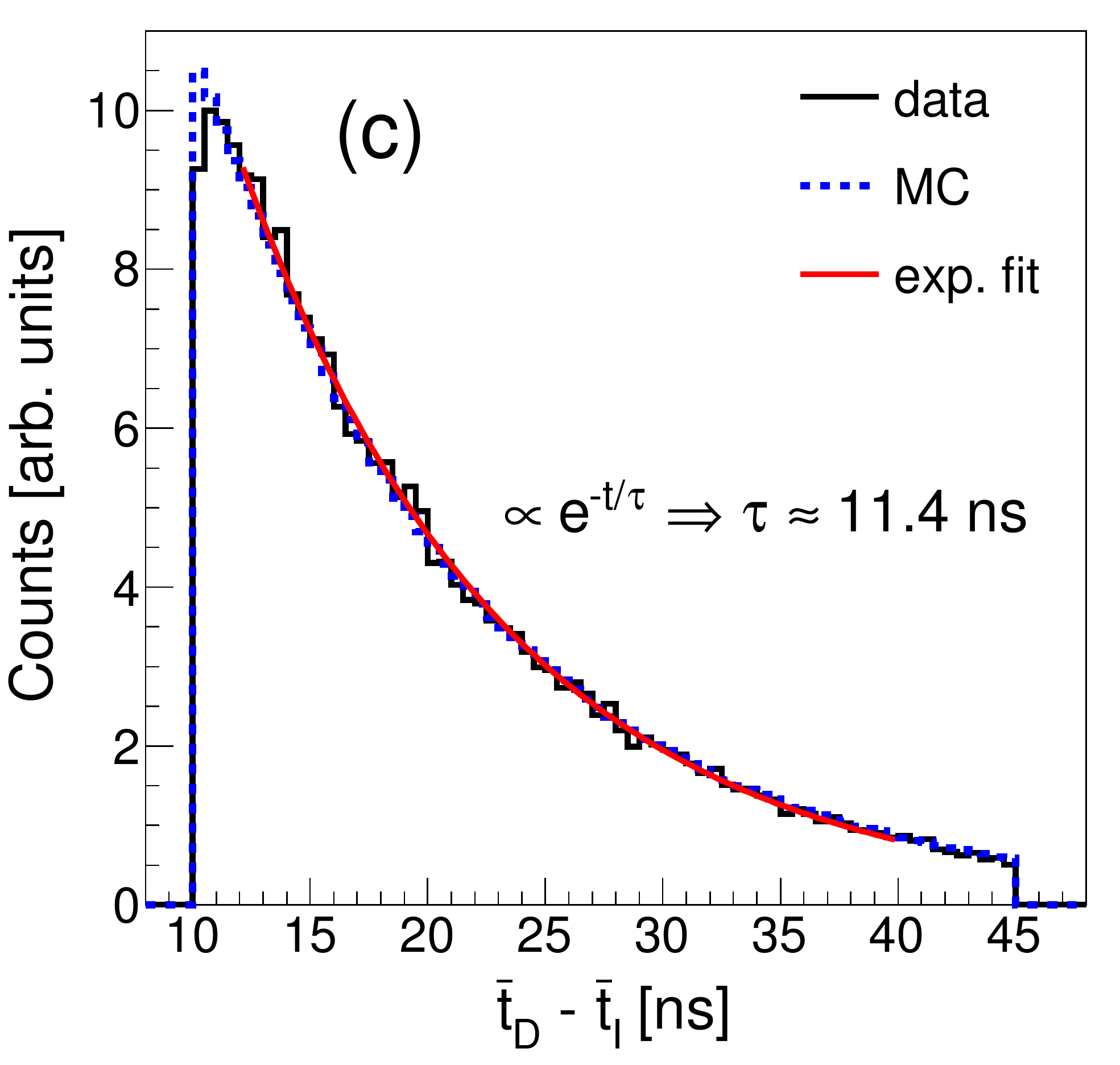}
\caption{$K^+$ detection in the Crystal 
Ball \cite{Werthmueller_NSTAR15_Proc}:
(a) Energy distribution of the $K^+$ decay clusters 
of all events (black solid), pionic decay
events (blue dotted) and muonic decay events
(red dashed).
(b) Comparison of experimental (black solid) and simulated
(blue dotted) decay cluster energy distributions for muonic decay
events.
(c) Impact-decay cluster time difference:
Experimental (black solid) and simulated (blue dotted)
distributions. Exponential fit to experimental
data (red curve).}
\label{fig:kaon_id}
\end{figure}

Fig.~\ref{fig:kaon_id} shows some characteristic
distributions related to the $K^+$-detection technique.
In (a), the decompositon of the total decay cluster
energy distribution (black histogram)
into the contributions
of the muonic (blue histogram) and the pionic 
(red histogram) decays is illustrated. A clear peak around
150 MeV due to the energy deposited by the $\mu^+$
can be seen, whereas the distribution coming from 
the pionic decay is broader. Fig.~\ref{fig:kaon_id}(b)
shows the good agreement of the experimental
and the simulated distributions in case of the muonic decay.
Only an overall energy correction is
necessary to account for the systematic shift.
Finally, the time difference between the impact
and the decay cluster is plotted in Fig.~\ref{fig:kaon_id}(c)
for experimental and simulated data. As expected, both 
distributions follow an exponential drop-off and
the decay time extracted from the experimental data
is in good agreement with the mean lifetime of the $K^+$-meson
\cite{Werthmueller_NSTAR15_Proc}.

\afterpage{%
\begin{figure}[!h]
\centering
\includegraphics[width=\textwidth]{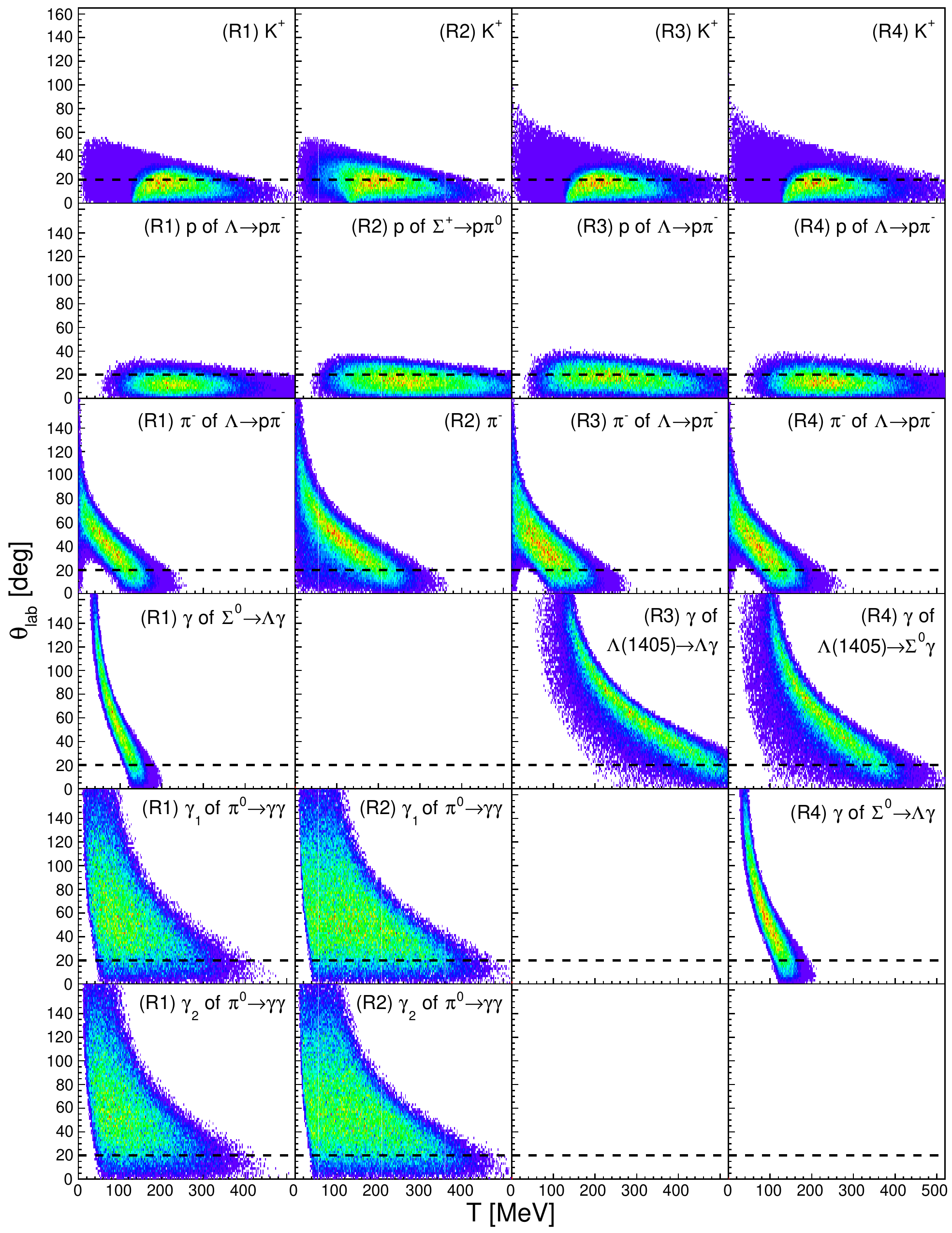}
\caption{Kinematics of $\gamma p\to K^+\Lambda(1405)$
in a simulation of the proposed A2 experiment.
\LF decays in columns, particles in rows:
(R1) \LF $\to\Sigma^0\pi^0$,
(R2) \LF $\to\Sigma^+\pi^-$,
(R3) \LF $\to\Lambda\gamma$,
(R4) \LF $\to\Sigma^0\gamma$
. The dashed line represents the CB-TAPS transition.}
\label{fig:kin}
\end{figure}
\clearpage
}

\subsection{Kinematics and event reconstruction}
\label{sec:kin_rec}
The kinematics and event topologies 
of the signal reactions (R1)--(R4)
were studied with simulations based on a Geant4 \cite{Geant4}
model of the experimental setup.
Events were generated according
to the differential cross sections measured by CLAS
\cite{Moriya_CS} and using the \LF resonance parameters 
provided by the PDG \cite{PDG15}. The resulting lab polar
angles and kinetic energies of all final state particles
for the proposed A2 experiment are shown in Fig.~\ref{fig:kin}.
For all reactions, most of the kaons will be going to TAPS
with kinetic energies mostly below the punch-through limit, so
that a clean detection via the in-crystal decay technique is
possible. The kinematics is similar for protons coming from the
$\Lambda$-decay. Time-of-flight, $dE/E$ and the BaF$_2$
pulse-shape analysis techniques will further help to achieve a clean 
identification of those particles in TAPS. The pions originating from the
decay of the $\Lambda$ are mainly going to CB. Their energies
are in the region were a clean separation from protons via $dE/E$
is possible. This will be more difficult for pions in TAPS since
their energies are higher and the $dE/E$ resolution is worse using
the TAPS vetos. Alternatively, one could use the 
still to be tested pizza detector. The MWPC could be used to reconstruct
the pion tracks but as the majority of protons will be detected in
TAPS, where no tracking is available, the exact determination of
the $\Lambda$-decay vertex will not be possible anyway.
Regarding the photons coming from the decays of the hyperons and the $\pi^0$, 
there will be no issues at all. Most of them are going to CB, where
they will contribute to the energy sum trigger.

The event reconstruction is requiring the detection of the complete
final state. The final analysis will exploit all possibilities for 
particle identification provided by the different detectors. In the
preliminary analyses presented here, the following selection criteria 
were applied:
\begin{itemize}
\item detection of $K^+$ cluster candidate via the subcluster finder
      algorithm, hit in PID or TAPS veto required. Only the 
      $K^+\to\mu^+\nu_\mu$ is considered in this test analysis.
\item identification of $\pi^-$ and proton via $dE/E$ in CB (PID) 
      and TAPS (vetos)
\item identification of photons via PID and veto detectors (no hit requested)
\end{itemize}
A kinematic fit \cite{Blobel_Kinfit} of the corresponding signal
reaction was performed on all particle
combinations fulfilling these criteria and the solution yielding the
lowest $\chi^2$ was used in the subsequent analysis steps. Depending
on the signal reaction, further cuts were
applied to obtain the results shown in the
following:
\begin{itemize}
\item $3\sigma$-cut (CL $< 2.7 \times 10^{-3}$) on the confidence 
      level of the best kinematic fit
\item $3\sigma$-cut on $m(\gamma\gamma)$ to select $\pi^0$-mesons or to exclude them
\item $3\sigma$-cut on the rest frame energy $E_\gamma^{\mathrm{rest}}$ of 
      the $\Sigma^0$ decay photon
\item $3\sigma$-cuts on the differences of the detected and 
      calculated proton polar and azimuthal angles
\end{itemize}
Furthermore, the simulated events were subject to a realistic model
of the A2 trigger consisting of an energy sum threshold for the Crystal
Ball detector and a multiplicity condition of logical units in CB
and TAPS. The units in CB are made from 45 sectors each 
containing up to 16 neighboring crystals, while 6 triangular sectors
are constituting the units in TAPS. In both detectors, one crystal per
unit exceeding the corresponding threshold will mark the unit as hit.
A condition is then applied on the number of hit units which roughly
corresponds to the number of detected particle clusters (not taking into
account multiple clusters in a single unit or a single cluster
spreading over several units).

\begin{figure}[!t]
\centering
\includegraphics[width=\textwidth]{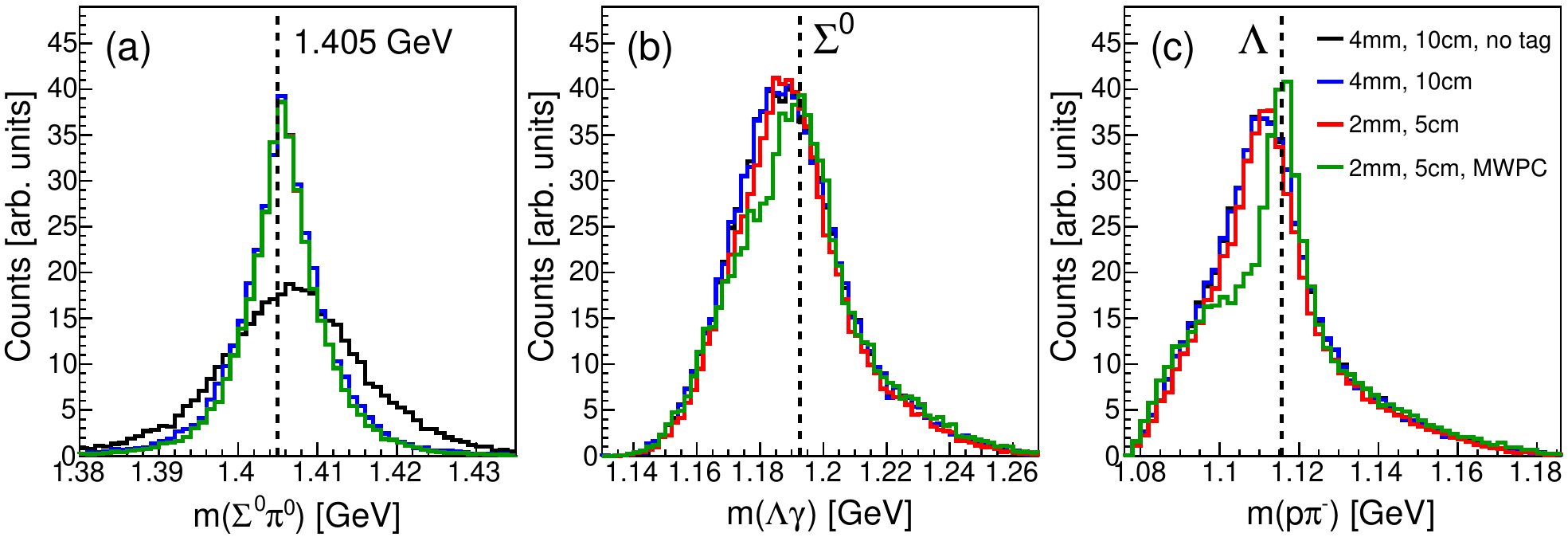}
\caption{Resolutions at $m(\Sigma\pi) = 1.405$ GeV 
in the $\Lambda$(1405)$\to\Sigma^0\pi^0$
analysis: 
(a) final $\Sigma^0\pi^0$ invariant mass showing the \LF resolution.
(b) $\Lambda\gamma$ invariant mass showing the $\Sigma^0$ signal.
(c) $p\pi^-$ invariant mass showing the $\Lambda$ signal.
The colors of the histograms denote different experimental
conditions and event reconstruction techniques (see text).
}
\label{fig:res_sigma0pi0}
\end{figure}

\begin{figure}[!t]
\centering
\includegraphics[width=0.68\textwidth]{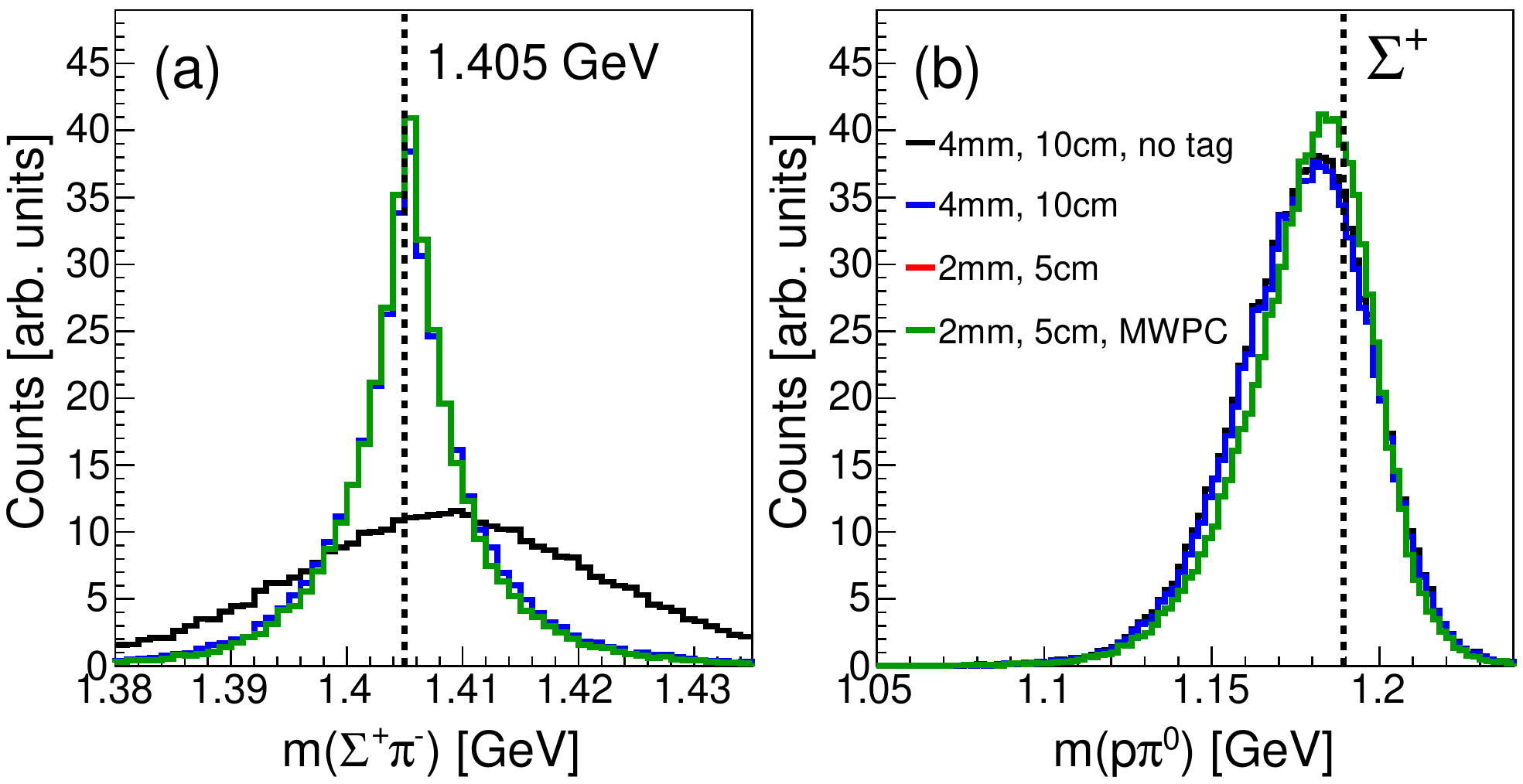}
\caption{Resolutions at $m(\Sigma\pi) = 1.405$ GeV 
in the $\Lambda$(1405)$\to\Sigma^+\pi^-$
analysis: 
(a) final $\Sigma^+\pi^-$ invariant mass showing the \LF resolution.
(b) $p\pi^0$ invariant mass showing the $\Sigma^+$ signal.
The colors of the histograms denote different experimental
conditions and event reconstruction techniques (see text).
}
\label{fig:res_sigmappim}
\end{figure}

The final resolutions of the $\Sigma^0\pi^0$ and $\Sigma^+\pi^-$ invariant 
mass distributions in the region of interest are shown in
Figs.~\ref{fig:res_sigma0pi0}(a) and \ref{fig:res_sigmappim}(a). They were
obtained by simulating events with $m(\Sigma\pi) = 1.405$ GeV
(zero intrinsic width). Two
different experimental scenarios were considered. In the first case,
a 4 mm photon beam collimator and a 10 cm long target cell was used. This 
corresponds to the experimental setup of the 2014 $\eta'$-experiment.
In the second case, a 2 mm collimator and a 5 cm target was tested. The
motivation for these settings was mainly to investigate if a spatially
more restricted primary vertex would improve the final $m(\Sigma\pi)$
resolutions. This seems not to be the case. Furthermore, the effect
of using the MWPC for the track reconstruction of the charged particles
was explored. Only a minor improvement in the $m(\Sigma\pi)$ distributions
can be observed. Approximated Gaussian FWHM resolutions in both the
$\Sigma^0\pi^0$ and $\Sigma^+\pi^-$ invariant masses of
10 MeV are observed. Finally, the scenario of running an untagged experiment,
i.e., without using the information of the photon tagger, was tested.
The advantage of running untagged would be potentially higher beam currents.
But, since in this case the beam photon energy has an uncertainty of 
\mbox{$\sim$78 MeV},
the resulting $\Sigma^0\pi^0$ and $\Sigma^+\pi^-$ resolutions are
twice and three times worse, respectively, than when running with the tagger.

Figs.~\ref{fig:res_sigma0pi0}(b),(c) and \ref{fig:res_sigmappim}(b) show
the signal distributions of the different intermediate hyperons of the
two $\Sigma\pi$ analysis channels. Again, only a minor improvement
of the more restricted primary vertex setup can be seen. The use of the
MWPC gives the most notable improvement in terms of signal width and
position for the $\Lambda$-reconstruction, which then improves also
slightly the $\Sigma^0$-signal. On the other hand, the reconstruction
of the more short-lived $\Sigma^+$ seems not to profit significantly.
As the calculations of the shown hyperon distributions are not using
the photon tagger information, the untagged scenario does not yield 
different distributions.

\subsection{Detection efficiencies and background contamination}
\label{sec:deteff_bgcont}
The detection efficiencies of reactions (R1)--(R4) were determined
with simulations of the standard experimental setup using a 4 mm 
photon beam collimator and a 10 cm long target cell (equivalent
to the 2014 $\eta'$-experiment). The main questions to be addressed
in this section is the optimal choice of the experimental trigger
and the amount of background contamination in all analyses channels.

Different thresholds in the range of 100--550 MeV for the energy sum
trigger in CB were investigated. A trigger multiplicity of 4 and higher
(M4+) was used along with all energy sum thresholds with the exception
of the 550 MeV setting, in which no multiplicity condition was applied
to reproduce the trigger of the 2014 $\eta'$-experiment. The M4+ multiplicity
condition denotes the highest reasonable multiplicity sensitive for 
reactions (R1)--(R4), where at least four particles need to be detected,
while rejecting large parts of events having less than four final state
particles. The use of M4+ is motivated by the findings discussed in the 
following. Namely, it was found that a low CB energy sum threshold
increases the detection efficiencies for all reactions of interest. 
On the other hand, a low threshold will let pass a large number of unwanted
events with higher production cross sections, possibly saturating the
data acquisition system. Therefore, the multiplicity condition should
be as restrictive as possible. As most particles of reactions (R1)--(R4)
going to TAPS have kinetic energies above 80 MeV, the threshold for the
multiplicity sectors in TAPS were set to this value to simulate
optimized settings concerning background rejection. The threshold for the CB
units were left at the standard value of 30 MeV.

Different sources of possible background contributions were checked in the
analyses of the signal reactions (R1)--(R4). They can be divided 
into four groups that will be briefly discussed in the next paragraphs.
More specific comments concerning the background contamination in
each of the signal reactions will be given afterwards.

\afterpage{%
\begin{figure}[!t]
\centering
\includegraphics[width=\textwidth]{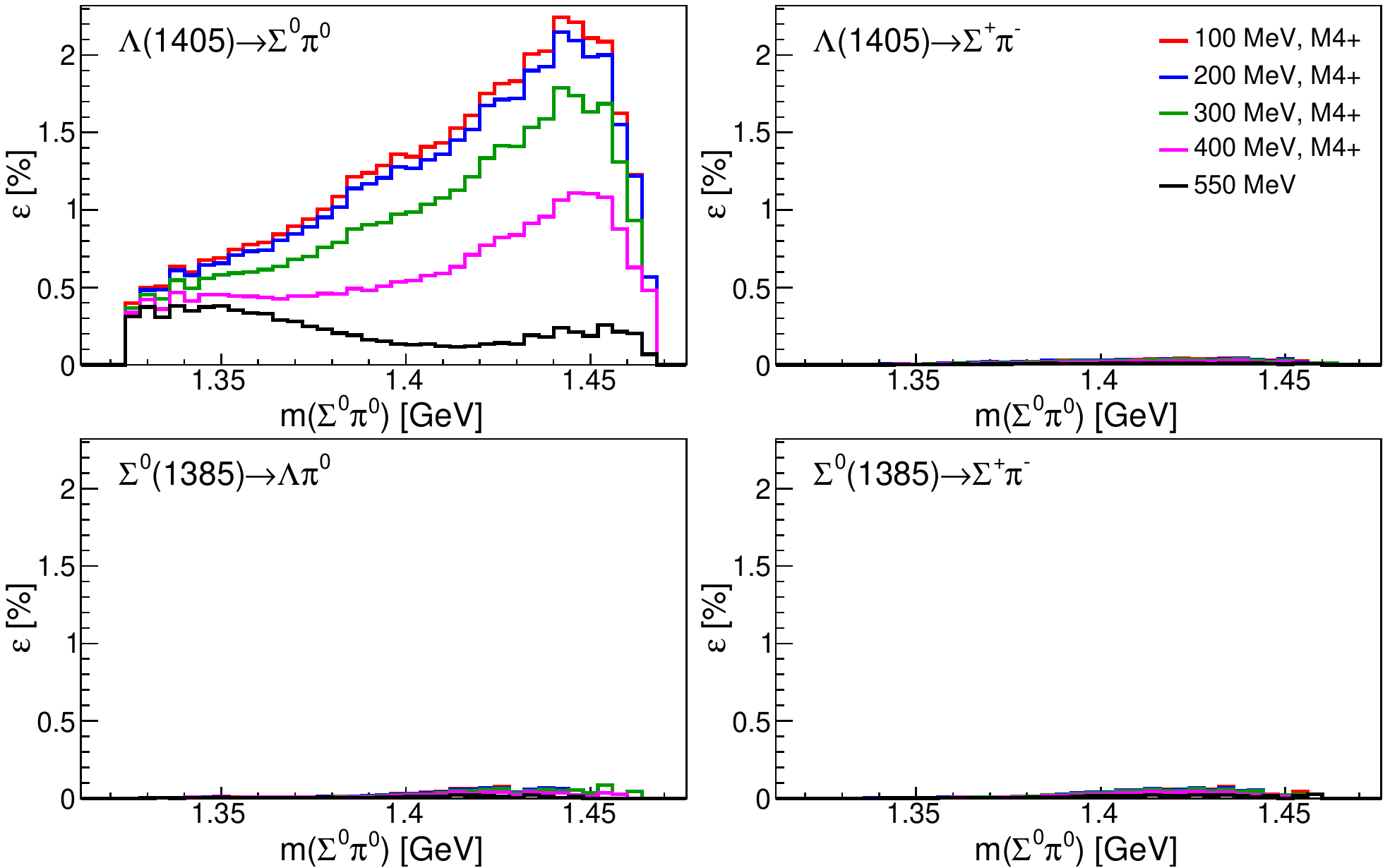}
\caption{Detection efficiencies of signal
(top left) and $\Lambda$(1405)/$\Sigma^0$(1385) hadronic
decay background contributions
in the 
$\Lambda$(1405)$\to\Sigma^0\pi^0$
analysis as a function of the $\Sigma^0\pi^0$
invariant mass.
}
\label{fig:eff_sigma0pi0}
\end{figure}
\begin{figure}[!t]
\centering
\includegraphics[width=\textwidth]{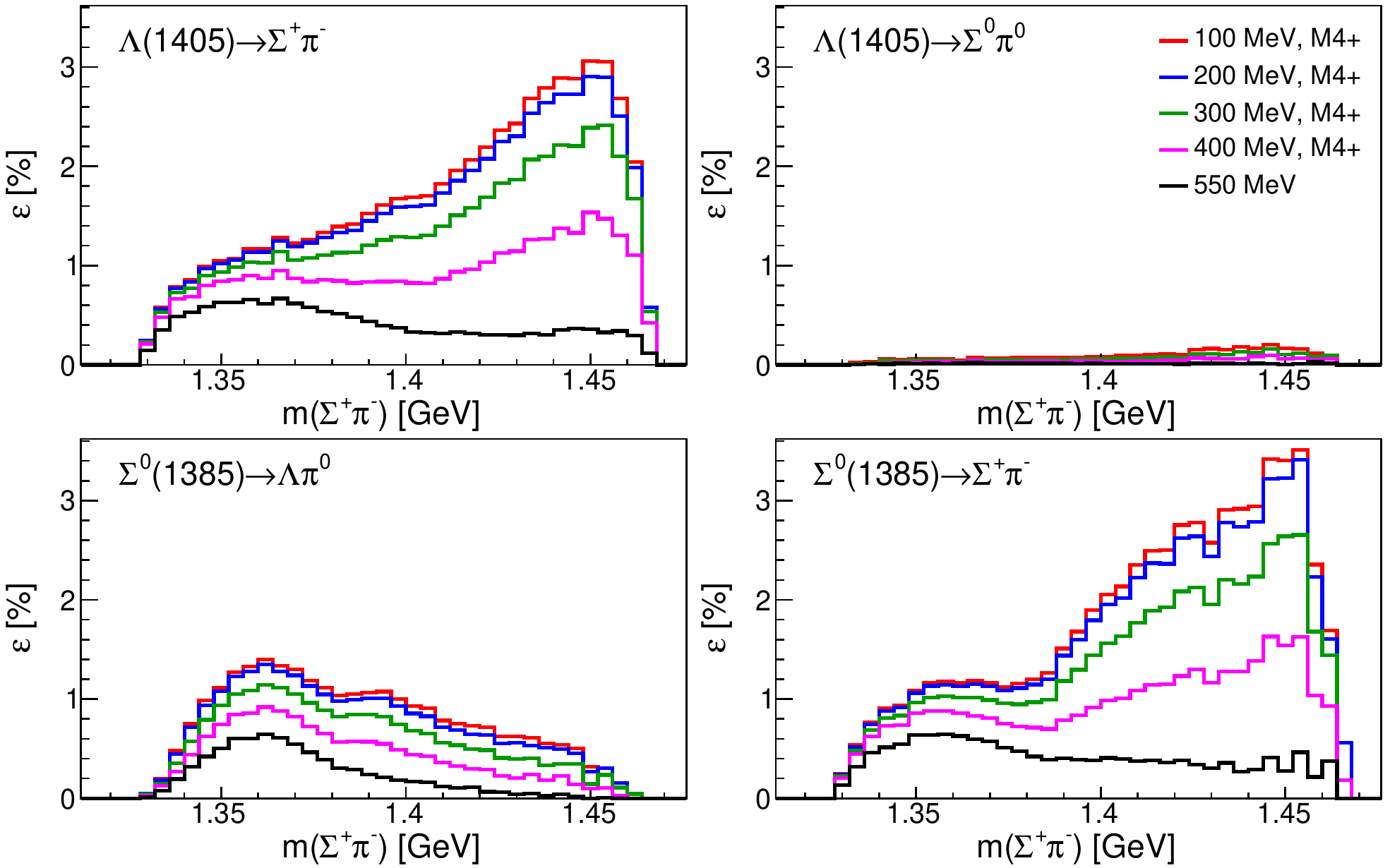}
\caption{Detection efficiencies of signal
(top left) and $\Lambda$(1405)/$\Sigma^0$(1385) hadronic
decay background contributions
in the 
$\Lambda$(1405)$\to\Sigma^+\pi^-$
analysis as a function of the $\Sigma^+\pi^-$
invariant mass.
}
\label{fig:eff_sigmappim}
\end{figure}
\clearpage
}

\paragraph{Nonstrange background}
Nonstrange reactions, such as
$\gamma p\to p\pi^+\pi^-\pi^0$ (also resonant 
via $\eta / \omega$-photoproduction),
$\gamma p\to p\eta\pi^0$,
and
$\gamma p\to p\omega\pi^0$ with 
$\eta / \omega\to\pi^+\pi^-\pi^0$
could mock the signal final states because $\pi^\pm$ (and 
also neutrons)
can be marked as kaons by the in-crystal decay detection algorithm.
The probability for this to happen is very small so that in general
the $K^+$-tag given by the algorithm provides a very characteristic
signature for strangeness photoproduction events.
Nevertheless, as the nonstrange 
cross sections are much higher than the one for the $\Lambda$(1405)-production,
possible contamination from these channels was investigated. It was found
that only a negligible amount of events pass the particle number and type
selection criteria, hence it is expected that other reactions
yielding the same (or different by $\pm$1 photon) final states are
also very unlikely to contribute significantly. The application
of the further analysis cuts discussed in Sec.~\ref{sec:kin_rec}
completely removed all nonstrange background events in all four analyses.

\paragraph{Ground-state hyperon production}
The reactions $\gamma p\to K^+\Lambda$ and
$\gamma p\to K^+\Sigma^0$ have roughly 10 times higher
cross sections than $\gamma p\to K^+\Lambda$(1405) and
could contaminate the signal reactions due to the presence of the kaon.
Especially reaction (R3) could be affected since it has exactly the same
final state as $\Sigma^0$-production. Indeed, a notable contamination was found
in the analysis of simulated data but since the \LF and the $\Sigma^0$ decay
photons have different energies, the removal of this background channel is
straightforward and can be done by a simple cut. No contamination was found
for the other reactions.

\paragraph{$\boldsymbol{\Sigma^0}$(1385)-production}
The cross section of $\gamma p\to K^+\Sigma^0$(1385)
is almost 3 times higher than the $\Lambda$(1405)-production
cross section in the region covered by this experiment.
In addition, some of the decays (see Tab.~\ref{tab:hyperon_decays})
lead to the same final states as in the \LF signal reactions. While (R1) is
not affected in first order due to $\Sigma^0(1385)\not\to\Sigma^0\pi^0$
(isospin forbidden), (R2) will see some contributions from
$\Sigma^0(1385)\to\Sigma^+\pi^-$. As the radiative decays of the
$\Sigma^0$(1385) and the \LF yield exactly the same final states,
contamination is inevitable for (R3) and (R4) and a simultaneous
analysis for the two hyperons needs to be performed.

\paragraph{Mutual contamination of signal reactions}
Because of the equal or similar final states of the signal
reactions (R1)--(R4), also mutual contamination is possible
and needs to be investigated. Contamination of the hadronic
$\Lambda$(1405)-decays by the radiative decays can of course be neglected due to the
tiny branching ratios of the latter, while contamination into the
other direction could be substantial. 
\\ \\
The detection efficiencies of signal and background reactions 
in the $\Lambda$(1405)$\to\Sigma^0\pi^0$ analysis (R1)
are plotted in Fig.~\ref{fig:eff_sigma0pi0} as a function of
the $\Sigma^0\pi^0$ invariant mass. As mentioned before, different
CB energy sum trigger thresholds were simulated and the corresponding
efficiencies are shown by curves of different color. The average signal 
efficiencies (top left) range from 0.23\% (550 MeV sum trigger) to
1.25\% (100 MeV, M4+). The factor of $\sim$5.4 between these two 
extreme cases illustrates the strong influence of the energy sum trigger, 
especially for higher values of $m(\Sigma^0\pi^0)$, where most kaons are going to
TAPS and cannot contribute to the deposited energy in CB. For lower values
of $m(\Sigma^0\pi^0)$ the efficiency is generally lower because there
are more high-energy undetected kaons punching-through the detectors.
As shown in the other plots of Fig.~\ref{fig:eff_sigma0pi0}, the efficiency
for the considered background channels is only a few percent with respect
to the signal efficiency. Nevertheless, considering all factors a
contamination of $\sim$17\% coming from $\Sigma^0$(1385)$\to\Lambda\pi^0$
must be expected in the worst case (see Tab.~\ref{tab:background}).

\afterpage{%
\begin{figure}[!t]
\centering
\includegraphics[width=\textwidth]{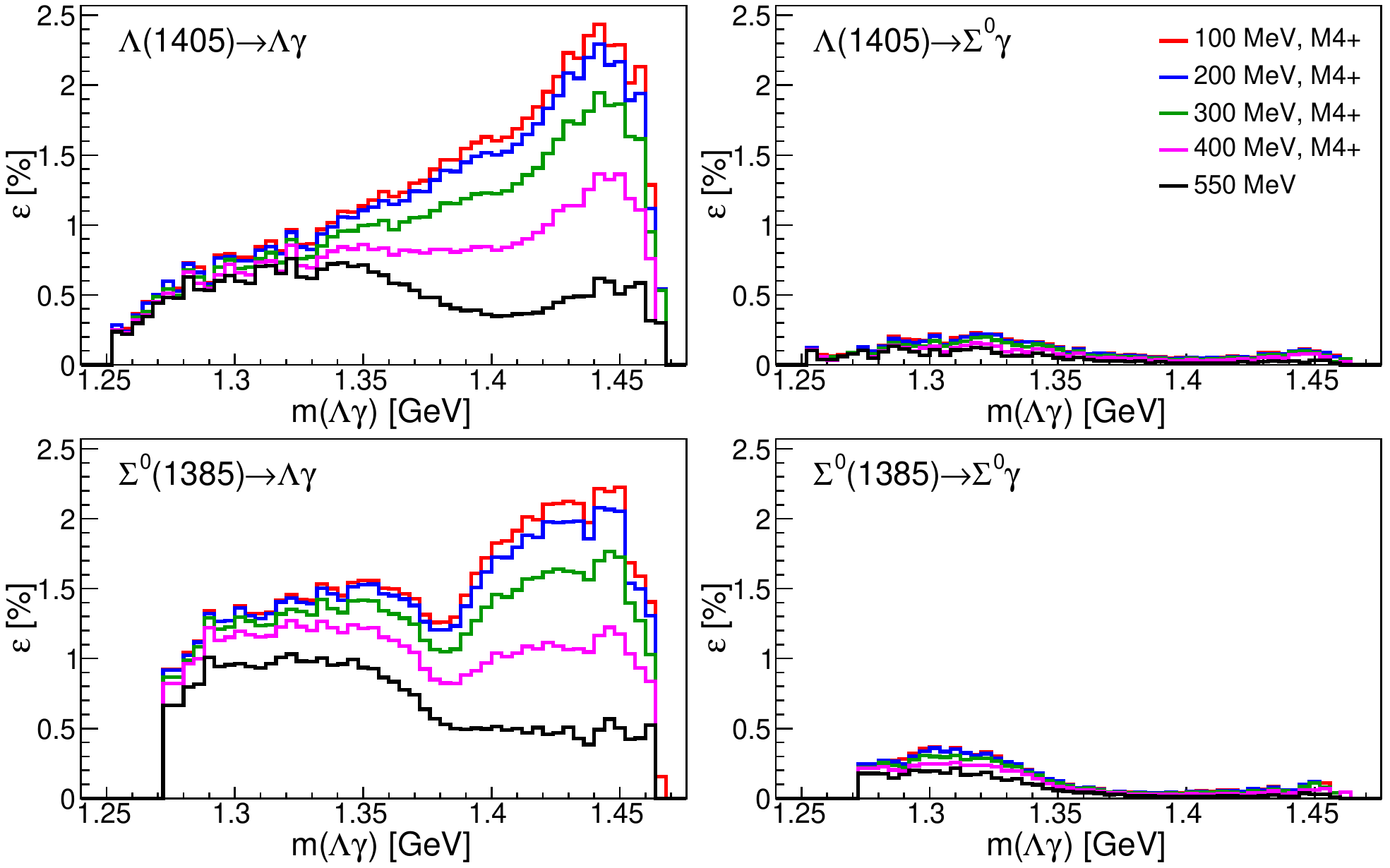}
\caption{Detection efficiencies of signal
(top left) and
$\Lambda$(1405)/$\Sigma^0$(1385) radiative decay
background contributions in the 
$\Lambda$(1405)$\to\Lambda\gamma$
analysis as a function of the $\Lambda\gamma$
invariant mass.
}
\label{fig:eff_radL}
\end{figure}
\begin{figure}[!t]
\centering
\includegraphics[width=\textwidth]{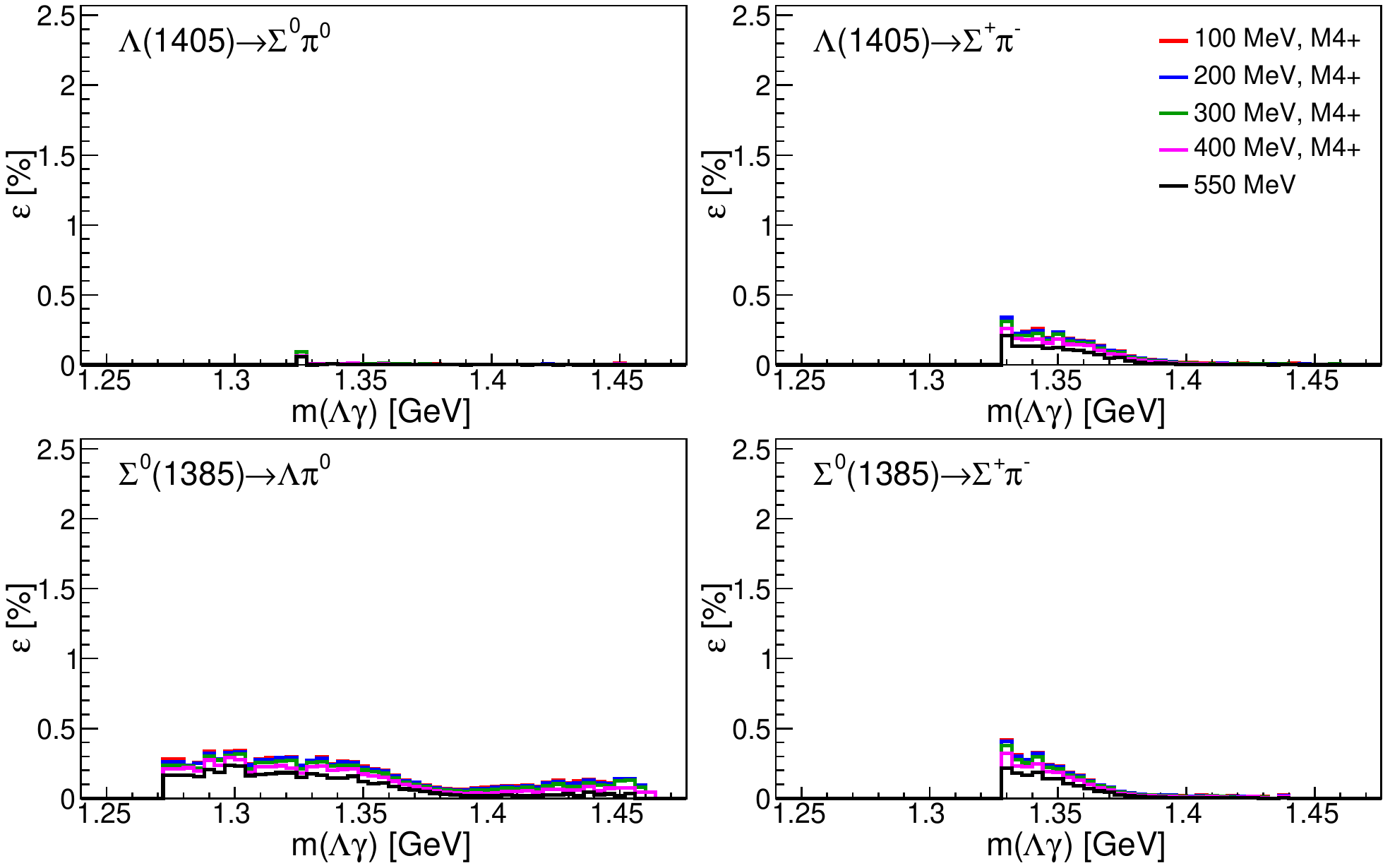}
\caption{Detection efficiencies of 
the $\Lambda$(1405)/$\Sigma^0$(1385) hadronic decay
background contributions in the 
$\Lambda$(1405)$\to\Lambda\gamma$
analysis as a function of the $\Lambda\gamma$
invariant mass.
}
\label{fig:eff_radL_2}
\end{figure}
\clearpage
}

\afterpage{%
\begin{figure}[!t]
\centering
\includegraphics[width=\textwidth]{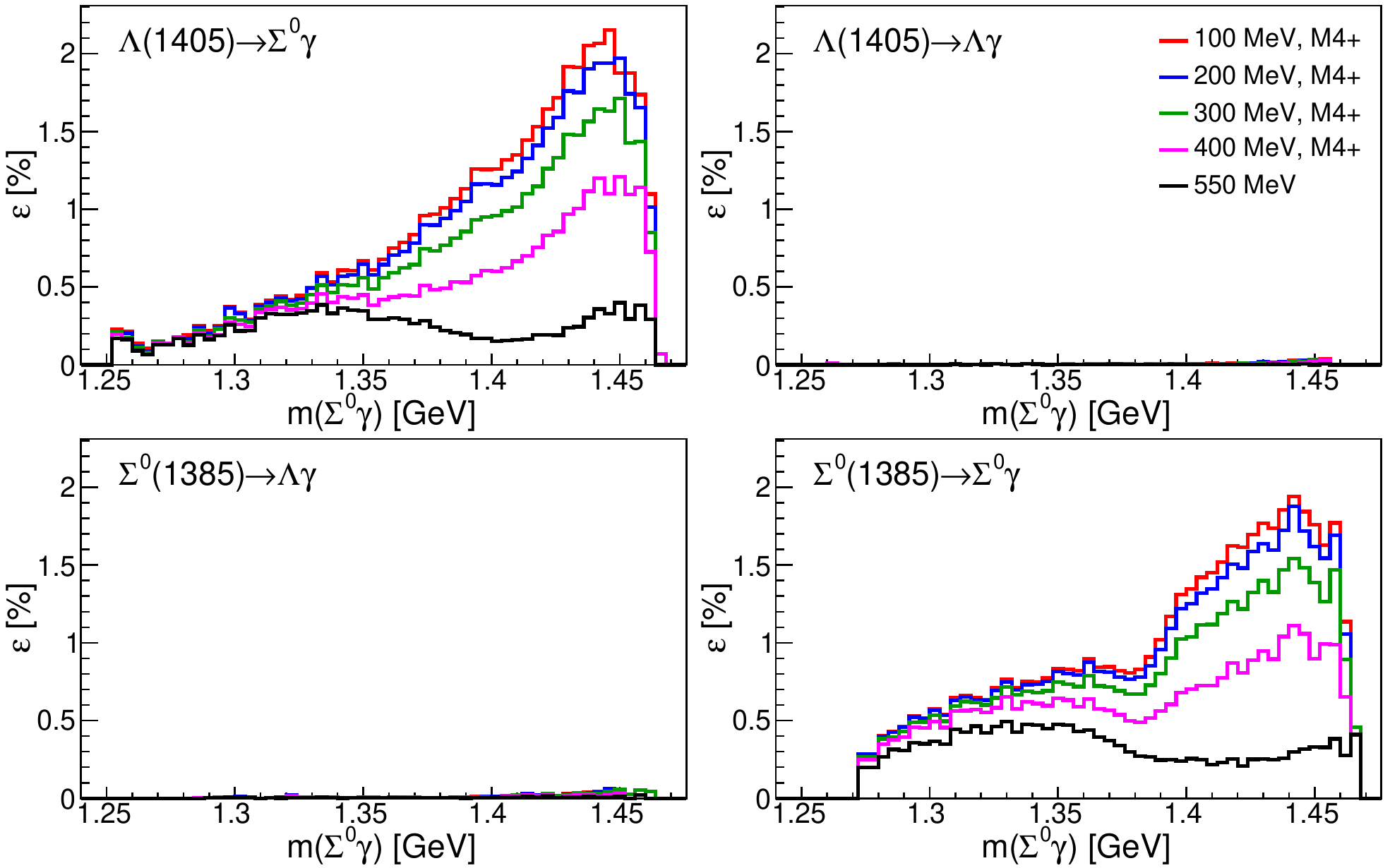}
\caption{Detection efficiencies of signal
(top left) and
$\Lambda$(1405)/$\Sigma^0$(1385) radiative decay
background contributions in the 
$\Lambda$(1405)$\to\Sigma^0\gamma$
analysis as a function of the $\Sigma^0\gamma$
invariant mass.
}
\label{fig:eff_radS}
\end{figure}
\begin{figure}[!t]
\centering
\includegraphics[width=\textwidth]{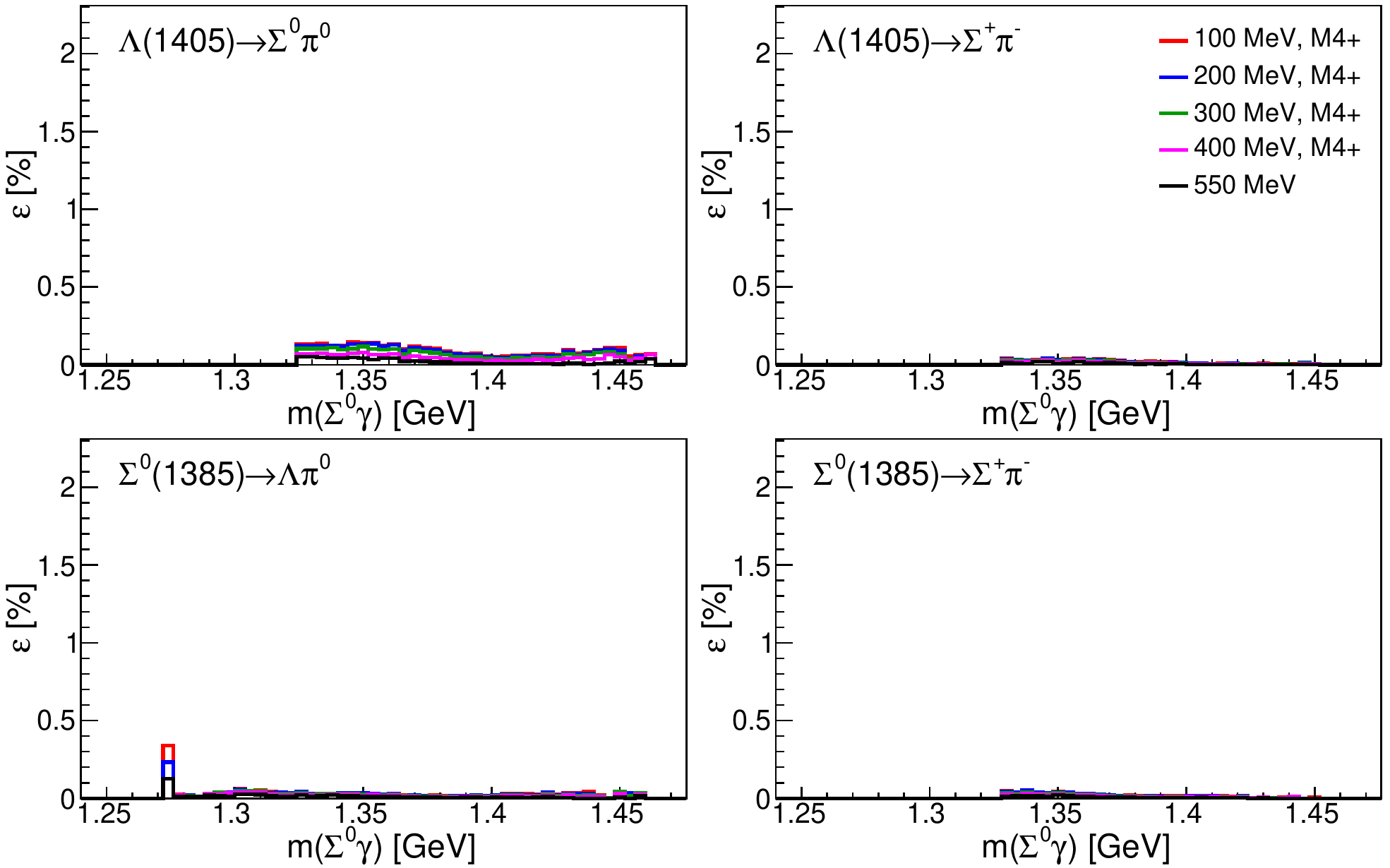}
\caption{Detection efficiencies of 
the $\Lambda$(1405)/$\Sigma^0$(1385) hadronic decay
background contributions in the 
$\Lambda$(1405)$\to\Sigma^0\gamma$
analysis as a function of the $\Sigma^0\gamma$
invariant mass.
}
\label{fig:eff_radS_2}
\end{figure}
\clearpage
}

The detection efficiencies of signal and background reactions 
in the $\Lambda$(1405)$\to\Sigma^+\pi^-$ analysis (R2)
are plotted in Fig.~\ref{fig:eff_sigmappim} as a function of
the $\Sigma^+\pi^-$ invariant mass. The average signal 
efficiencies (top left) range from 0.41\% (550 MeV sum trigger) to
1.70\% (100 MeV, M4+) and are thus significantly higher than for (R1).
This is probably mostly due to the lower multiplicity final state
and the less distant secondary vertex of the $\Sigma^+$-hyperon.
The efficiency of the $\Sigma^0$(1385)$\to\Sigma^+\pi^-$ background
reaction is even slightly higher than for the signal reaction, while
the efficiency of $\Sigma^0$(1385)$\to\Lambda\pi^0$, which leads to
the exactly same final state, is considerably lower. The efficiency of
the $\Lambda$(1405)$\to\Sigma^0\pi^0$ decay is very small. Taken
into account all factors (see Tab.~\ref{tab:background}), the
$\Sigma^0$(1385)$\to\Lambda\pi^0$ seems to give a major contribution
roughly more than four times stronger than the signal. The simple analysis
performed for this work is therefore not sufficient for the signal
extraction and some better methods need to be applied in the final
analysis. Once the background contributions will have been suppressed
sufficiently, a separation of the different channels should be 
possible for example by fitting the different $m(\Sigma^+\pi^-)$
distributions (see Fig.~\ref{fig:ls_sigmappim}) to the experimental
spectrum.

The detection efficiencies of signal and background reactions 
in the $\Lambda$(1405)$\to\Lambda\gamma$ analysis (R3)
are plotted in Figs.~\ref{fig:eff_radL} and \ref{fig:eff_radL_2}
as a function of the $\Lambda\gamma$ invariant mass. Here possible
contamination from all hadronic (with the exception of $\Sigma^-\pi^+$)
and radiative decays of the \LF
and the $\Sigma^0$(1385) were considered. The average signal 
efficiencies (top left of Fig.~\ref{fig:eff_radL}) range 
from 0.51\% (550 MeV sum trigger) to
1.25\% (100 MeV, M4+). Due to the high energetic \LF decay 
photon contributing to the CB energy sum, the efficiency for low
values of $m(\Lambda\gamma)$, where also more kaons are detected in
CB, is almost the same for all energy sum thresholds.
Obviously the
$\Sigma^0$(1385)$\to\Lambda\gamma$ decay has a equally high
efficiency, while the radiative decays of both excited hyperons
to the $\Sigma^0\gamma$ final state are much small although still
notable. On the same low level are the contaminations from the hadronic decays 
shown in Fig.~\ref{fig:eff_radL_2} with the exception of the
$\Lambda$(1405)$\to\Sigma^0\pi^0$ channel, which is even more
suppressed. According to Tab.~\ref{tab:background}
the $\Sigma^0$(1385)$\to\Lambda\pi^0$ will be the dominant background
in this analysis but basically all final states except 
$\Lambda$(1405)$\to\Sigma^0\gamma$ will contaminate the signal. It has
to be stressed again that the performed analysis was rather simple and
only exploited a limited number of cuts. Surely
the different distributions of all contributions in, e.g., the 
$\Lambda\gamma$ invariant mass (Fig.~\ref{fig:ls_radL}) or the
\LF decay photon energy in the \LF rest frame (Fig.~\ref{fig:ls_radL_2})
will be useful for disentangling the various channels in a more elaborate
analysis.

The detection efficiencies of signal and background reactions 
in the $\Lambda$(1405)$\to\Sigma^0\gamma$ analysis (R4)
are plotted in Figs.~\ref{fig:eff_radS} and \ref{fig:eff_radS_2}
as a function of the $\Sigma^0\gamma$ invariant mass. As before, possible
contamination from all hadronic and radiative decays of the \LF
and the $\Sigma^0$(1385) were considered. The average signal 
efficiencies (top left of Fig.~\ref{fig:eff_radS}) range 
from 0.25\% (550 MeV sum trigger) to
0.87\% (100 MeV, M4+).
Again, the corresponding radiative decay of the $\Sigma^0$(1385)
has an equally high efficiency. The other radiative decays only have very low
efficiencies.
Efficiencies of the hadronic decays of both hyperons
are low but because of the high production cross section and branching
ratio, the $\Sigma^0$(1385)$\to\Lambda\pi^0$ decay will
be the dominant background again (see Tab.~\ref{tab:background}).
The separation of this and all other backgrounds will make use
of the different $\Lambda\gamma$ invariant mass and \LF decay photon energy 
distributions shown in Fig.~\ref{fig:ls_radS} and
Fig.~\ref{fig:ls_radS_2}, respectively.

\begin{table}[!t]
\centering
\begin{tabular}{@{}lccccr@{}}
\toprule
& 
$\epsilon_{\mathrm{det}}$ [\%] & 
$\Gamma_i / \Gamma\, (Y^*)$ &
$\Gamma_i / \Gamma\, (Y)$ &
$\sigma$ [nb] & 
\multicolumn{1}{c}{$c_{\mathrm{eff}}$} \\
\midrule
{\bf (R1)} $\boldsymbol{\Lambda}${\bf(1405)}$\boldsymbol{\to\Sigma^0\pi^0}$ \\[0.05mm]
signal
&  1.19  &  33\% &  64\% &  154 &  1.00 \\
$\Lambda$(1405)$\to\Sigma^+\pi^-$
&  0.02  &  33\% &  52\% &  154 &  0.01 \\
$\Sigma^0$(1385)$\to\Lambda\pi^0$
&  0.03  &  87\% &  64\% &  404 &  0.17 \\
$\Sigma^0$(1385)$\to\Sigma^+\pi^-$
&  0.03  &   6\% &  52\% &  404 &  0.01 \\
\\[-3mm]
{\bf (R2)} $\boldsymbol{\Lambda}${\bf(1405)}$\boldsymbol{\to\Sigma^+\pi^-}$ \\[0.05mm]
signal
&  1.62  &  33\% &  52\% &  154 &  1.00 \\
$\Lambda$(1405)$\to\Sigma^0\pi^0$ 
&  0.07  &  33\% &  64\% &  154 &  0.05 \\
$\Sigma^0$(1385)$\to\Lambda\pi^0$
&  0.74  &  87\% &  64\% &  404 &  3.89 \\
$\Sigma^0$(1385)$\to\Sigma^+\pi^-$
&  1.73  &   6\% &  52\% &  404 &  0.51 \\
\\[-3mm]
{\bf (R3)} $\boldsymbol{\Lambda}${\bf(1405)}$\boldsymbol{\to\Lambda\gamma}$ \\[0.05mm]
signal
&  1.18  &  $1.39 \times 10^{-3}$ &  64\% &  154 &  1.00 \\
$\Lambda$(1405)$\to\Sigma^0\gamma$
&  0.11  &  $1.39 \times 10^{-3}$ &  64\% &  154 &  0.09 \\
$\Sigma^0$(1385)$\to\Lambda\gamma$
&  1.50  &  $1.25 \times 10^{-2}$ &  64\% &  404 &  29.99 \\
$\Sigma^0$(1385)$\to\Sigma^0\gamma$
&  0.15  &  $1.25 \times 10^{-2}$ &  64\% &  404 &  3.00 \\
$\Lambda$(1405)$\to\Sigma^0\pi^0$
&  0.01  &  33\% &  64\% &  154 &  2.01 \\
$\Lambda$(1405)$\to\Sigma^+\pi^-$
&  0.08  &  33\% &  52\% &  154 &  13.08 \\
$\Sigma^0$(1385)$\to\Lambda\pi^0$
&  0.17  &  87\% &  64\% &  404 &  236.55 \\
$\Sigma^0$(1385)$\to\Sigma^+\pi^-$
&  0.10  &   6\% &  52\% &  404 &  7.80 \\
\\[-3mm]
{\bf (R4)} $\boldsymbol{\Lambda}${\bf(1405)}$\boldsymbol{\to\Sigma^0\gamma}$ \\[0.05mm]
signal
&  0.82  &  $1.39 \times 10^{-3}$ &  64\% &  154 &  1.00 \\
$\Lambda$(1405)$\to\Lambda\gamma$
&  0.01  &  $1.39 \times 10^{-3}$ &  64\% &  154 &  0.01 \\
$\Sigma^0$(1385)$\to\Lambda\gamma$
&  0.02  &  $1.25 \times 10^{-2}$ &  64\% &  404 &  0.58 \\
$\Sigma^0$(1385)$\to\Sigma^0\gamma$
&  0.96  &  $1.25 \times 10^{-2}$ &  64\% &  404 &  27.62 \\
$\Lambda$(1405)$\to\Sigma^0\pi^0$
&  0.09  &  33\% &  64\% &  154 &  26.06 \\
$\Lambda$(1405)$\to\Sigma^+\pi^-$
&  0.02  &  33\% &  52\% &  154 &  4.70 \\
$\Sigma^0$(1385)$\to\Lambda\pi^0$
&  0.03  &  87\% &  64\% &  404 &  60.07 \\
$\Sigma^0$(1385)$\to\Sigma^+\pi^-$
&  0.02  &   6\% &  52\% &  404 &  2.24 \\
\bottomrule
\end{tabular}
\caption{Estimation of effective contributions $c_{\mathrm{eff}}$ 
for the 200 MeV CB energy sum, M4+ trigger scenario based on detection 
efficiencies $\epsilon_{\mathrm{det}}$, branching ratios $\Gamma_i / \Gamma$
for the excited hyperon \mbox{$Y^* = \{\Lambda(1405), \Sigma^0(1385)\}$} and the secondary
hyperon \mbox{$Y = \{\Lambda, \Sigma^0\}$}, and the production cross sections $\sigma$.}
\label{tab:background}
\end{table}

\subsection{Extraction of observables and cross checks}
An estimation of the to be expected effective strengths of the signal and background 
contributions is shown in Tab.~\ref{tab:background}. The detection efficiencies
correspond to the 200 MeV CB energy sum, M4+ trigger scenario. The branching ratios
for the hadronic decays were taken from Tab.~\ref{tab:hyperon_decays}. 
The radiative decay widths suffer from considerable uncertainties caused
by the large range of theoretical values and the absence
of a previous direct measurement (see Sec.~\ref{sec:intro_raddec}).
We motivate our choice by the most recent results from unitary chiral theory
(see Eqns.~\ref{unit_chi_pol_res_lg} and \ref{unit_chi_pol_res_sg}) and
use an average value for both decays of
$\Gamma_{Y\gamma} = 70$ keV for all further calculations. This is also close
to the overall average of the model calculations shown
in Tab.\ I of Ref.~\cite{Taylor:2005bn}.
The assumed branching ratios are thus
$\Gamma_{\Lambda\gamma} / \Gamma = \Gamma_{\Sigma^0\gamma} / \Gamma = 1.39 \times 10^{-3}$.
Also, the unknown $\Gamma_{\Sigma^0\gamma} / \Gamma$ for the $\Sigma^0$(1385)
was assumed to be equal to 
$\Gamma_{\Lambda\gamma} / \Gamma = 1.25 \times 10^{-2}$.
The production cross sections for the \LF and the $\Sigma^0$(1385) were
estimated from the CLAS measurements \cite{Moriya_CS} by linear interpolation from
the corresponding reaction threshold to the first data point at $E_\gamma = 1662$ MeV.
An average cross section in the photon beam energy range covered by the
endpoint tagger was then calculated. Finally, the effective contributions were
normalized to the signal contribution.

In all analysis channels, the $\Sigma^0$(1385)$\to\Lambda\pi^0$ 
contribution seems to be the largest background. As can be seen 
in Figs.~\ref{fig:ls_sigmappim}--\ref{fig:ls_radS_2}, a separation
using the shown variables (amongst others) should be possible with techniques
such as sPlot \cite{Pivk:2005ca}, but as the
magnitude of the background is rather large, a better rejection
needs to be implemented beforehand using more sophisticated analysis methods. 
For example, an optimization of the signal-to-background ratios
based on the confidence levels of kinematic fits of all reactions
candidates could be implemented. This approach was already successfully
used in the measurement of the $\Sigma^0$(1385)$\to\Lambda\gamma$ decay
width \cite{Keller:2011br}.

The following list summarizes several cross checks that can be
performed in the proposed experiment, and some additional advantages:
\begin{itemize}
\item The second $\Lambda$-decay $\Lambda\to n\pi^0$ provides an independent
      data set (with lower statistics) with a completely different final state
      and detection efficiencies. 
\item The radiative decays of the $\Sigma^0$(1385) will have to be extracted
      in parallel and the obtained result for $\Gamma_{\Lambda\gamma}$ can be
      compared to the previously measured value \cite{Taylor:2005bn,Keller:2011br}.  
\item The flux normalization can be checked via the determination of the 
	  photoproduction cross sections for the $\Sigma^0$(1385) or nonstrange
	  mesons.
\item Absolute normalization is only needed for the $(\Sigma\pi)^0$ line-shape
      observables.
\item Systematic uncertainties due to detection efficiencies will be
      small in the extraction of the beam-helicity asymmetry $I^\odot$ and
      the radiative decay widths because of cancellation effects.
\end{itemize}

\subsection{Running conditions}
\label{sec:exp_run_cond}
Due to the small $\Lambda$(1405)-production cross sections, the proposed
experiment should run at the highest possible luminosity. This includes
the use of a 4mm collimator and a 10 cm long hydrogen target. As discussed
in Sec.~\ref{sec:kin_rec}, the benefits of a smaller beam spot and a shorter
target would only be marginal, but on the other hand lead to a decrease
in luminosity by $\sim$75\%. Even higher luminosities would be possible
in an untagged experiment but this would reduce the photon beam energy 
resolution dramatically, which would lead to unacceptable resolutions
in the observables and to severe difficulties in the background rejection.
Unfortunately, high luminosities make the use of the MWPC impossible 
due to rate limitations, but,
as shown in Sec.~\ref{sec:kin_rec}, no substantial advantage enabled
by their use could be observed in the event reconstruction.

Test measurements using a 4mm collimator and a 10 cm long hydrogen target
but varying other experimental parameters
were performed in July 2014 in preparation of the
$\eta'$-experiment. Those measurements allow 
a more accurate estimation of the parameters for the proposed
experiment. The detection efficiencies presented in Sec.~\ref{sec:deteff_bgcont}
suggest that a 200 MeV CB energy sum, M4+ trigger seems to be a reasonable trigger
setting. Test data using a 250 MeV CB energy sum threshold without 
multiplicity condition is available \cite{Elog_4879} 
and shows the following distribution
of multiplicities: M1 (10\%), M2 (30\%), M3 (40\%), M4+ (20\%). Hence, we
deduce that applying the M4+ multiplicity condition could reduce the trigger
rate by up to a factor of 5 when using a 200 MeV CB energy sum threshold.
This would allow to increase the beam current by a factor of up to 5 compared to
the test measurement yielding $5 \times 14$ nA $= 70$ nA at a data acquisition
livetime of $\sim$60\%. This is close
to the beam current used during production running
of the $\eta'$-experiment (60 nA) and still well below
the hardware limit of the endpoint tagger (110 nA). A further increase of
the beam current towards this limit is only possible when the trigger rate is
reduced even more. Requiring a hit in the endpoint tagger for a positive trigger
decision would be an option to be tested for the proposed experiment.
For the $\eta'$-decays experiment in 2014, this mechanism was found to be not useful
since at 30 nA, a trigger reduction of only 10\% could
be achieved \cite{DAQ_Wiki_July_14_24_Meeting}.
Nevertheless, this depends on the main trigger settings and
the benefit of having the EPT contributing to the trigger
decision is expected to be higher for lower CB energy sum thresholds.
Namely, a reduction of the tagged event ratio from 11\% to 5\% was
observed when the energy sum threshold was lowered from 550 MeV
to 250 MeV.
Other options enabling even higher beam currents
would be a threshold Cerenkov detector vetoing electrons
and pions, or an FPGA-based kaon trigger in TAPS sensitive to the 
in-crystal decays. 

Smaller changes to the standard experimental settings
that could further optimize the signal detection efficiencies and the
background suppression, would be checked in detail before the experiment.
They include the use of the pizza detector as $dE$ and time-of-flight
detector for TAPS,
the position of this detector and TAPS with respect to the target, the 
closing of the backward hole in CB with spare BaF$_2$ crystals
or thick plastic scintillator veto detectors, and the use of a 
cylindrical PID detector inside CB with a larger radius in place
of the unusable MWPC.

\subsection{Combination with other experiments}
Combined running with another experiment is possible by, e.g., changing
the trigger conditions on a regular basis during the experiment or by
trigger prescaling. This could even
be helpful for obtaining data with a less restrictive trigger 
conditions for calibration purposes. The proposed trigger 
for this experiment combined with the higher-lying tagged
photon energy range could make the $a_0(980)/f_0(980)$ scalar meson 
production experiment (LOI to the PAC 2013 \cite{LOI_Scalar_Meson})
possible on hydrogen. If this experiment would be performed using
a deuterium target, we could consider measuring a part of our
proposed experiment (radiative decays) on that target as well. 
Finally, certain decays
of the $\eta'$ and $\omega$-mesons studied in the ongoing analyses
of the 2014 endpoint-tagger data could benefit from additional
data provided by the complementary trigger of our experiment.

\begin{figure}[!b]
\centering
\includegraphics[width=9cm]{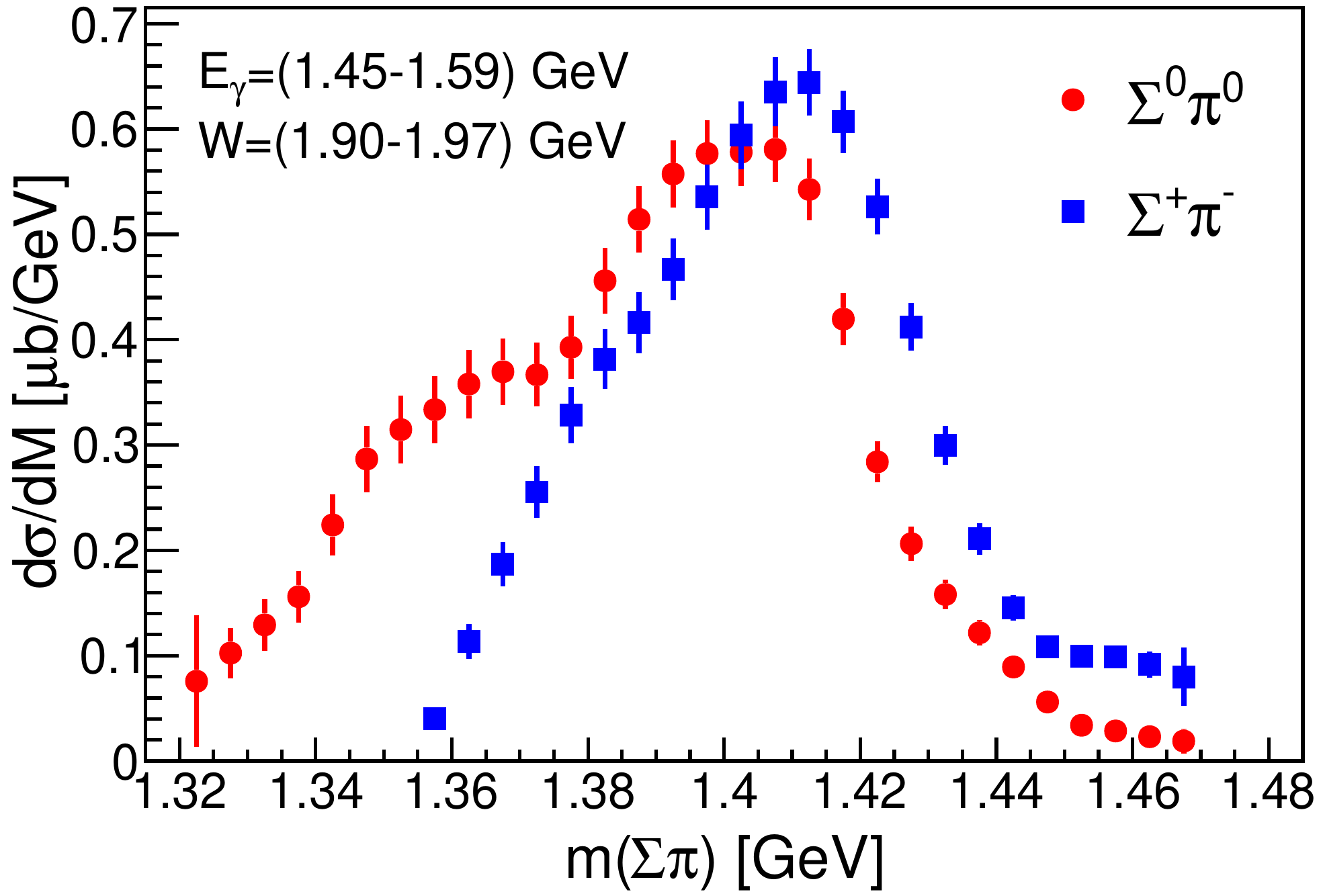}
\caption{Expected statistical quality of the \LF $(\Sigma\pi)^0$ line shapes
obtained in the proposed experiment.}
\label{fig:ls_A2}
\end{figure}

\clearpage

\section{Beamtime request}
The beamtime request is made based on the measurement of the 
$\Lambda$(1405)$\to\Sigma^0\gamma$ radiative decay width with 
a statistical uncertainty of 25\%. This corresponds to the 
uncertainty of the first
measurement of the $\Sigma^0$(1385)$\to\Lambda\gamma$ decay
width \cite{Taylor:2005bn}. Realistic numbers for the 
$\Lambda$(1405)$\to\Lambda\gamma$ will be very similar
after a more extended background subtraction, hence a
common calculation can be performed using the formula
\begin{equation}
\Delta t = \left[ \delta_{stat}^2 \cdot \sigma \cdot N_{e^-} \cdot
\epsilon_{tag} \cdot \rho_t \cdot \epsilon_{det} \cdot \epsilon_{daq} 
\cdot \Gamma_i/\Gamma \right] ^{-1}
\end{equation}
and the corresponding values for the tagged photon energy range 
$E_\gamma = $1450--1590 MeV
\\[2ex]
\begin{tabular}{@{}lll@{}}
$\sigma$          & production cross section    & 154 nb \\
$N_{e^-}$         & detected electrons          & $2.76 \times 10^7$ s$^{-1}$ at 70 nA\\
$\epsilon_{tag}$  & tagging efficiency          & 60\% \\
$\rho_t$          & target density              & 0.4215 b$^{-1}$ (10 cm LH$_2$) \\
$\epsilon_{det}$  & detection efficiency        & 0.90\% \\
$\epsilon_{daq}$  & DAQ livetime                & 60\% \\
$\Gamma_i/\Gamma$ & branching ratio             
& $1.39 \times 10^{-3} \cdot 64\% \cdot 84\%$ \\
$\delta_{stat}$   & relative statistical uncertainty & 25\% \\
\end{tabular}
\\[2ex]
where the additional branching ratio of 84\% comes
from the sum of the two analyzable $K^+$-decays 
($\Gamma_{\mu^+\nu_\mu}/\Gamma = 63.56\%$,
$\Gamma_{\pi^+\pi^0}/\Gamma = 20.67\%$) and $\epsilon_{det}$
is the weighted average detection efficiency for the two kaon-decay
analyses. In addition to the $\Delta t \approx 1000$ hours of production 
running, we ask for an additional block of 150 hours for setting up
and optimizing the experiment and request thus a total beamtime of
\begin{center}
\bf 1150 hours.
\end{center}
The corresponding projected results for the $(\Sigma\pi)^0$ line shapes
are shown in Fig.~\ref{fig:ls_A2}.

\section*{Acknowledgements}
D.W.\ acknowledges support by the Schweizerischer Nationalfonds 
(158822, 167759).

\clearpage

\appendix

\section{Additional figures}

\begin{figure}[!h]
\centering
\vspace{1cm}
\includegraphics[height=6.2cm]{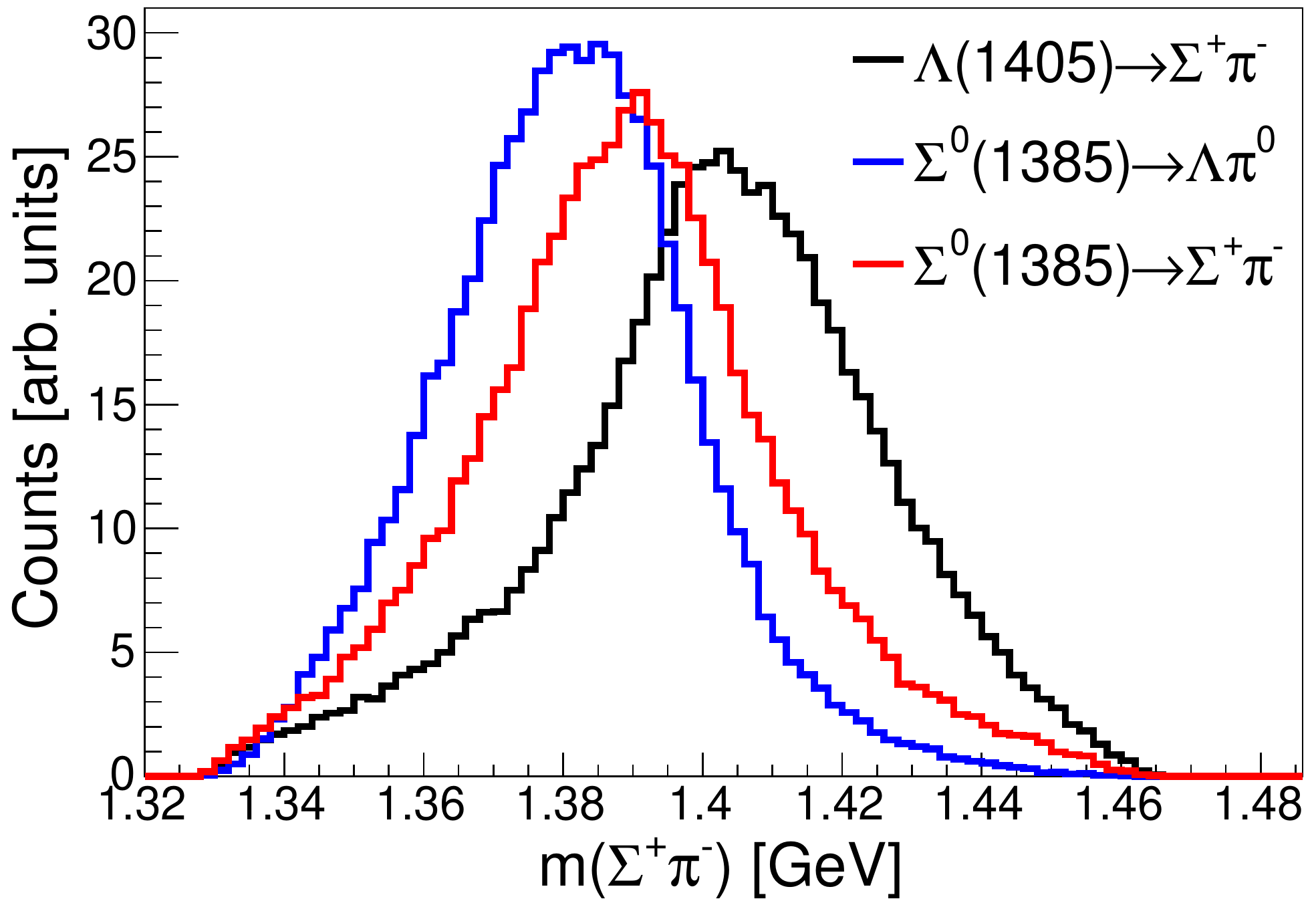}
\caption{$m(\Sigma^+\pi^-)$ distributions
(arb.\ scaling)
in the $\Lambda$(1405)$\to\Sigma^+\pi^-$
analysis for the signal (black curve) and the two
most important background contributions
(blue and red curves).
}
\label{fig:ls_sigmappim}
\end{figure}

\begin{figure}[!h]
\centering
\includegraphics[width=0.78\textwidth]{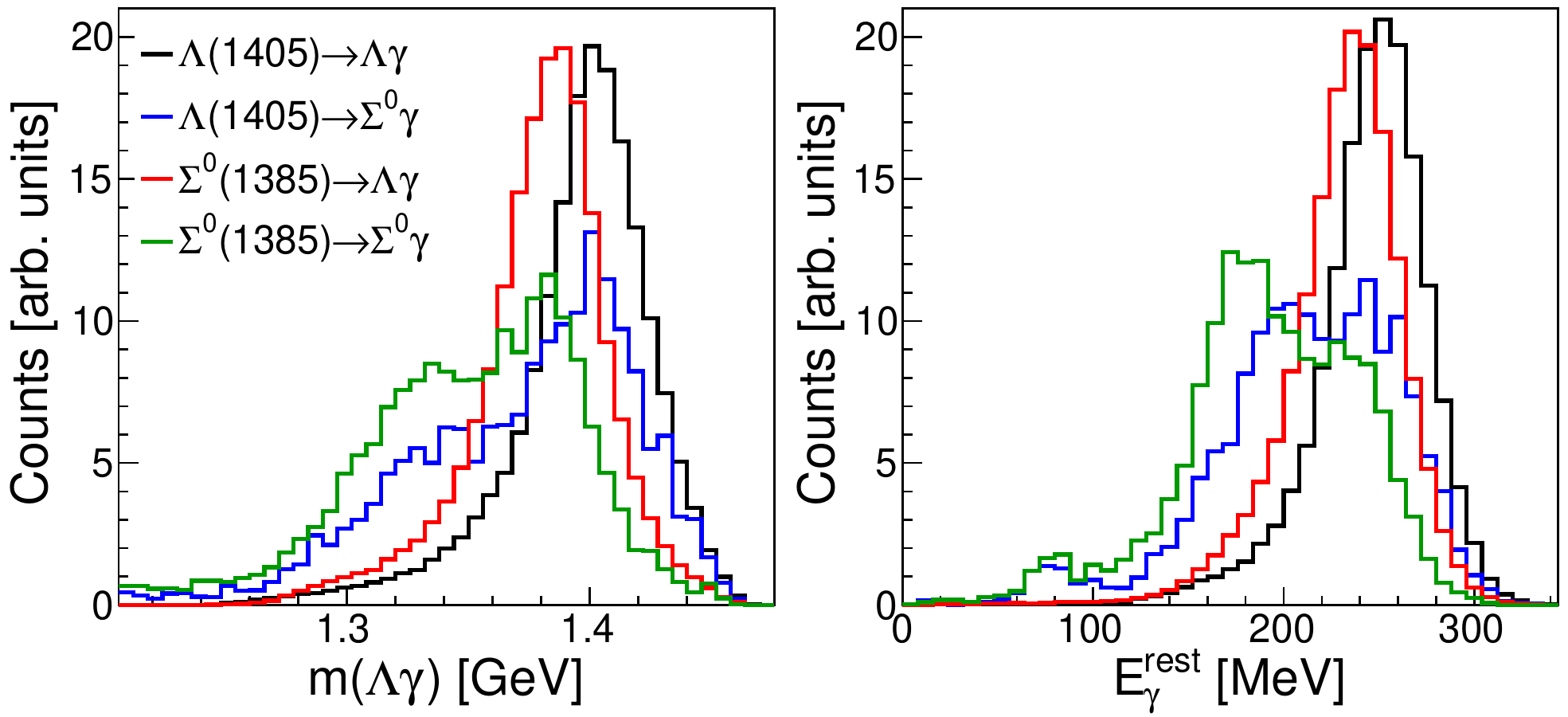}
\caption{Left side: $m(\Lambda\gamma)$ distributions
(arb.\ scaling)
in the $\Lambda$(1405)$\to\Lambda\gamma$
analysis for the signal (black curve) and the 
background caused
by the other $\Lambda$(1405)/$\Sigma^0$(1385) 
radiative decays (blue, red and
green curves). 
Right side: Corresponding distributions of the \LF decay photon energy
(\LF rest frame).
}
\label{fig:ls_radL}
\end{figure}

\begin{figure}[!h]
\centering
\includegraphics[width=0.78\textwidth]{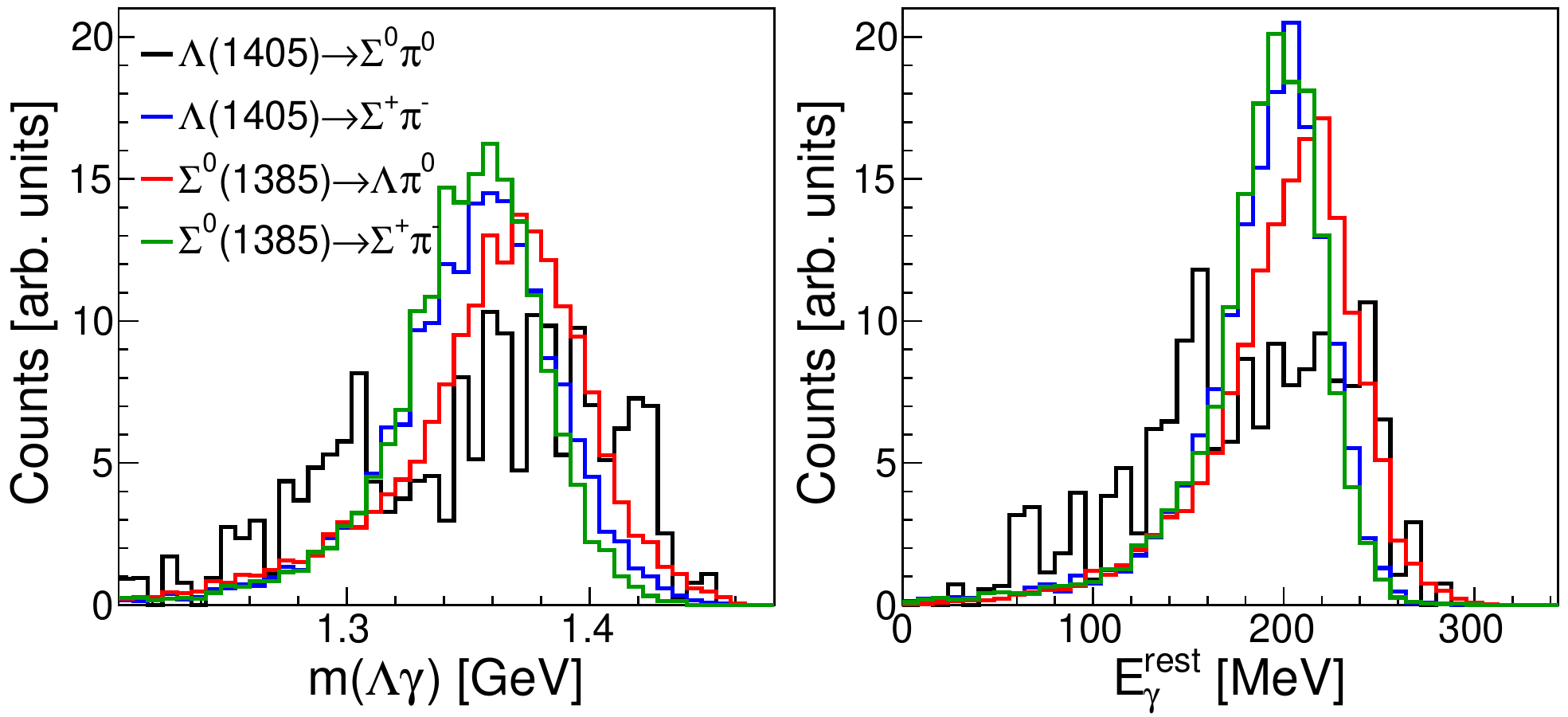}
\caption{Left side: $m(\Lambda\gamma)$ distributions
(arb.\ scaling)
in the $\Lambda$(1405)$\to\Lambda\gamma$
analysis for the 
background caused
by the $\Lambda$(1405)/$\Sigma^0$(1385) 
hadronic decays (blue, red and
green curves). 
Right side: Corresponding distributions of the \LF decay photon energy
(\LF rest frame).
}
\label{fig:ls_radL_2}
\end{figure}

\begin{figure}[!h]
\centering
\includegraphics[width=0.78\textwidth]{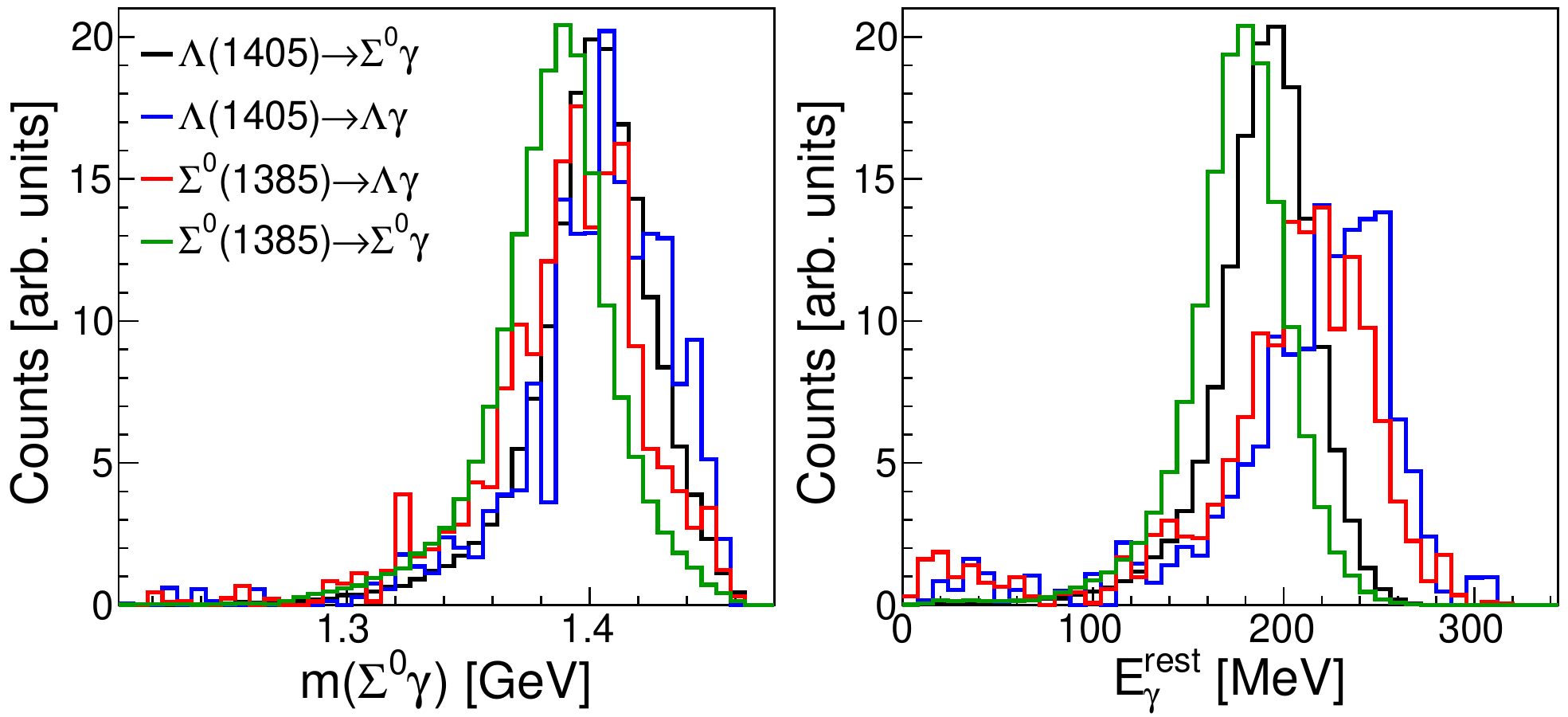}
\caption{Left side: $m(\Sigma^0\gamma)$ distributions
(arb.\ scaling)
in the $\Lambda$(1405)$\to\Sigma^0\gamma$
analysis for the signal (black curve) and the 
background caused
by the other $\Lambda$(1405)/$\Sigma^0$(1385) 
radiative decays (blue, red and
green curves). 
Right side: Corresponding distributions of the \LF decay photon energy
(\LF rest frame).
}
\label{fig:ls_radS}
\end{figure}

\begin{figure}[!h]
\centering
\includegraphics[width=0.78\textwidth]{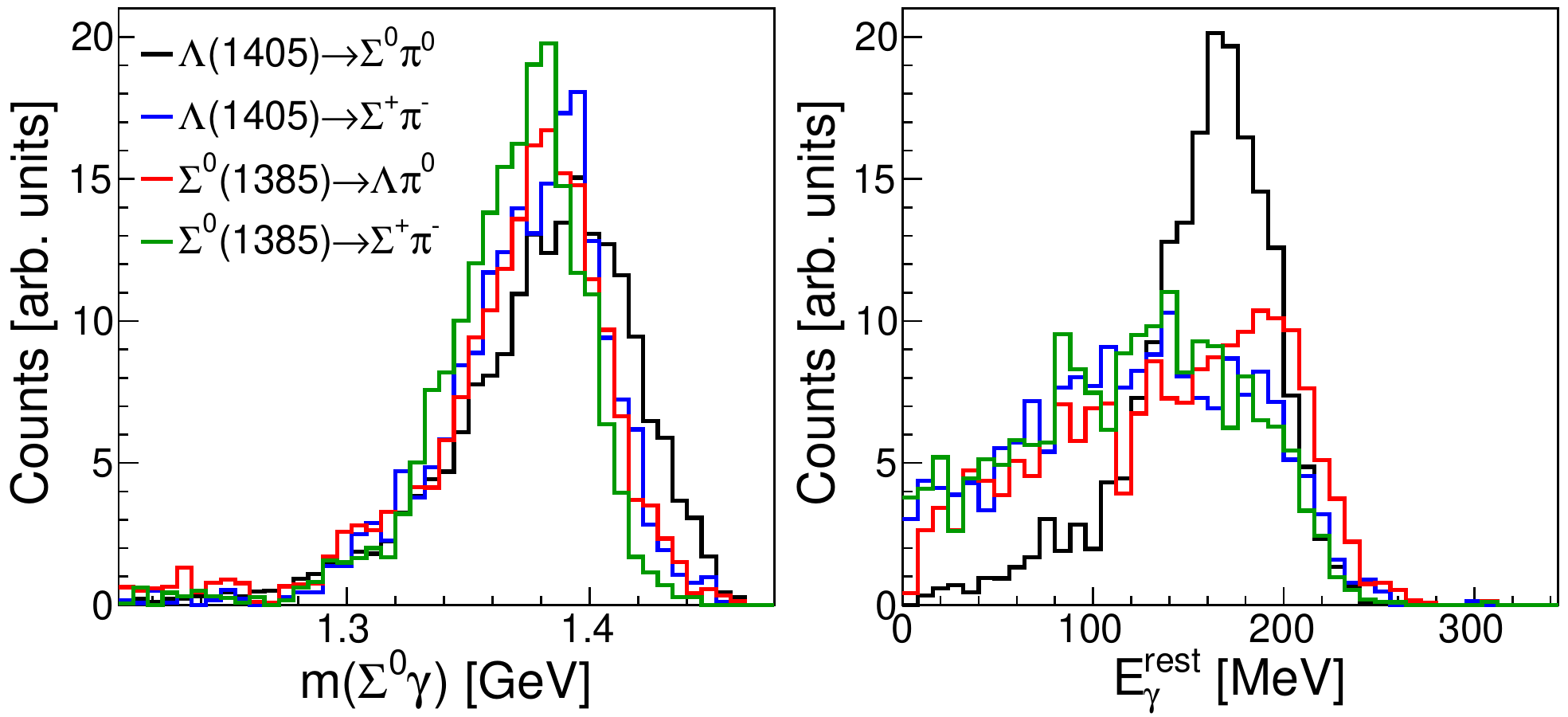}
\caption{Left side: $m(\Sigma^0\gamma)$ distributions
(arb.\ scaling)
in the $\Lambda$(1405)$\to\Sigma^0\gamma$
analysis for the 
background caused
by the $\Lambda$(1405)/$\Sigma^0$(1385) 
hadronic decays (blue, red and
green curves). 
Right side: Corresponding distributions of the \LF decay photon energy
(\LF rest frame).
}
\label{fig:ls_radS_2}
\end{figure}

\clearpage

\section*{References}
\printbibliography[heading=none,sorting=none]

\end{document}